\documentclass{article}
\usepackage{graphicx}
\graphicspath{{fig/}}
\usepackage{amsmath}
\usepackage{amssymb}
\usepackage{booktabs}
\usepackage[dvipsnames]{xcolor}
\usepackage{tikz}
\usepackage[a4paper, total={6in, 8in}]{geometry}
\usepackage{xspace}
\usepackage[style=numeric-comp,backend=biber,autocite=inline,natbib=true,sorting=none,sortcites=true]{biblatex}
\usepackage{rotating}
\usepackage{siunitx}
\usepackage{tablefootnote}
\usepackage{csquotes}
\usepackage{lineno}
\usepackage{authblk}
\usepackage{hyperref}
\hypersetup{
    colorlinks=false,
    linkcolor=blue,
    urlcolor=blue,
    citecolor=black
}

\usepackage{datetime}
\newcommand{\version}[1]{[Version {#1} - \today, \currenttime\ {\scriptsize GMT}]}
\def\currentVersion{\version{$\beta.1$}}
\usepackage[color=gray]{background}
\backgroundsetup{
    position={+13.5cm,-16cm},
    scale=1.4,
    firstpage=true, 
    angle=90,
    opacity=0.5,    
    contents=\currentVersion    
}

\newcommand{\pT}{p_{\perp}} 
\newcommand{\pTo}{p_{\perp 0}}
\newcommand{\ie}{\textit{i.e.}\ }
\newcommand{\eg}{\textit{e.g.}\ }

\newcommand{\n}{\newline}

\newcommand{\xmax}{\ensuremath{X_\text{max}}\xspace}
\newcommand{\xmumax}{\ensuremath{X_{\mu,\text{max}}}\xspace}
\newcommand{\nmu}{\ensuremath{N_\mu}\xspace}

\newcommand{\tablecaption}[1]{\caption{#1}\vspace{0.3em}}
\newcommand{\avg}[1]{\langle #1 \rangle}
\newcommand{\lna}{\ln\!A}
\newcommand{\mlna}{\avg\lna}
\newcommand{\slna}{\sigma(\lna)}
\newcommand{\sinel}{\sigma_\text{inel}}
\newcommand{\mult}{N_\text{mult}}
\newcommand{\inelast}{\frac{E_\text{max}}E}
\newcommand{\piefrac}{\frac{E_{\pi^0}}E}
\newcommand{\sigdb}{\sigma_{D,B}}
\newcommand{\rivet}{\textsc{Rivet}\xspace}
\newcommand{\pythia}{\textsc{Pythia}\xspace}
\newcommand{\corsika}{\textsc{Corsika}\xspace}
\newcommand{\apriori}{\textit{a priori}\xspace}
\newcommand{\sibyll}[1]{\mbox{\textsc{Sibyll}}#1\xspace}
\newcommand{\sqrtsnn}{\ensuremath{\sqrt{s_\text{NN}}}\xspace}
\newcommand{\dpmjet}[1]{\mbox{\textsc{DPMJet\ifthenelse{\equal{#1}{}}{}{-#1}}}\xspace}
\newcommand{\qgsjet}[1]{\mbox{\textsc{QGSJet#1}}\xspace}
\newcommand{\epos}{\textsc{EPOS}\xspace}
\newcommand{\eposlhc}{\textsc{EPOS\,LHC}\xspace}
\newcommand{\conex}{\textsc{Conex}\xspace}
\newcommand{\mceq}{{\sc MCEq}\xspace}
\newcommand{\crpropa}{\textsc{CRPropa}\xspace}
\newcommand{\seneca}{\textsc{Seneca}\xspace}
\newcommand{\fluka}{\textsc{Fluka}\xspace}
\newcommand{\proposal}{\textsc{Proposal}\xspace}

\newcommand{\ppColl}{p+p\xspace}
\newcommand{\pipColl}{$\pi$\kern0.1em-$p$\xspace}
\newcommand{\pPbColl}{p+Pb\xspace}
\newcommand{\piAColl}{$\pi+A$\xspace}
\newcommand{\pAColl}{p+$A$\xspace}
\newcommand{\pOColl}{p+O\xspace}

\newcommand{\pNColl}{p+N\xspace}
\newcommand{\AAColl}{$A+A$\xspace}
\newcommand{\PbPbColl}{Pb+Pb\xspace}
\newcommand{\pipmpColl}{$\pi^\pm$+p\xspace}

\newcommand{\pipCColl}{$\pi^+$+C\xspace}

\newcommand{\KpmpColl}{K$^\pm$+p\xspace}

\newcommand{\piz}{\pi^0}

\newcommand{\TppColl}{pp\xspace}
\newcommand{\TpipColl}{$\pi$\kern0.1em-$p$\xspace}
\newcommand{\TpPbColl}{pPb\xspace}

\newcommand{\TpAlColl}{pAl\xspace}
\newcommand{\TpCColl}{pC\xspace}
\newcommand{\TpBeColl}{pBe\xspace}

\newcommand{\TpipmpColl}{$\pi^\pm$p\xspace}
\newcommand{\TpippColl}{$\pi^+$p\xspace}

\newcommand{\TppbarColl}{p$\bar{{\rm p}}$\xspace}
\newcommand{\TpimpColl}{$\pi^-$p\xspace}
\newcommand{\TKppColl}{K$^+$p\xspace}
\newcommand{\TKpmpColl}{K$^\pm$p\xspace}
\newcommand{\TpimCColl}{$\pi^-$C\xspace}

\title{Road map for the tuning of hadronic interaction models with accelerator-based and astroparticle data }

\author[1,2,3,\footnote{Corresponding authors}]{J.~Albrecht}
\author[1,4,5]{J.~Becker Tjus}
\author[2]{N.~Behling}
\author[6]{J.~Blazek}
\author[7]{M.~Bleicher}
\author[2]{J.~Boelhauve}
\author[8]{L.~Cazon}
\author[9a,9b]{R.~Concei\c{c}\~ao}
\author[1,2]{H.~Dembinski}
\author[2]{L.~Dietrich}
\author[6]{J.~Ebr}
\author[2]{J.~Ellbracht}
\author[10]{R.~Engel}
\author[12]{A.~Fedynitch}
\author[13]{M.~Fieg}
\author[14]{M.V.~Garzelli}
\author[15,*]{C.~Gaudu}
\author[16]{G.~Graziani}
\author[2]{P.~Gutjahr}
\author[10]{A.~Haungs}
\author[10,11]{T.~Huege}
\author[2]{K.~Hymon}
\author[2]{M.~Hünnefeld}
\author[15,1,*]{K.-H.~Kampert}
\author[2]{L.~Kardum}
\author[2]{L.~Kolk}
\author[17]{N.~Korneeva}
\author[1,2]{K.~Kröninger}
\author[18]{A.~Maire}
\author[19]{H.~Menjo}
\author[15]{L.~Morejon}
\author[14]{S.~Ostapchenko}
\author[20, 20b]{P.~Paakkinen}
\author[10]{T.~Pierog}
\author[21]{P.~Plotko}
\author[12]{A.~Prosekin}
\author[21,22,23]{L.~Pyras}
\author[24]{T.~Pöschl}
\author[15]{J.~Rautenberg}
\author[10]{M.~Reininghaus}
\author[1,2,3]{W.~Rhode}
\author[1,2,*]{F.~Riehn}
\author[10]{M.~Roth}
\author[15]{A.~Sandrock}
\author[26]{I.~Sarcevic}
\author[27]{M.~Schmelling}
\author[14]{G.~Sigl}
\author[28]{T.~Sjöstrand}
\author[23]{D.~Soldin}
\author[10]{M.~Unger}
\author[20]{M.~Utheim}
\author[6]{J.~Vícha}
\author[29]{K.~Werner}
\author[2]{M.E.~Windau}
\author[30]{V.~Zhukov}

\affil[1]{Ruhr Astroparticle and Plasma Physics Center (RAPP Center), Bochum, Germany}
\affil[2]{Department of Physics, TU Dortmund University, D-44221 Dortmund, Germany}
\affil[3]{Lamarr Institute for Machine Learning and Artificial Intelligence, Dortmund, Germany}
\affil[4]{Theoretical Physics IV: Plasma Astroparticle Physics, Ruhr University Bochum, 44780 Bochum, Germany}
\affil[5]{Department of Space, Earth and Environment, Chalmers University of Technology, Gothenburg, Sweden}
\affil[6]{FZU -- Institute of Physics of the Czech Academy of Sciences, Prague, Czech Republic}
\affil[7]{Institute for Theoretical Physics, Goethe University Frankfurt, Frankfurt am Main, Germany}
\affil[8]{Instituto Galego de Física de Altas Enerxías (IGFAE), Universidade de Santiago de Compostela, Santiago de Compostela, Spain}
\affil[9a]{Physics Department, Instituto Superior T\'ecnico (IST), University of Lisbon, Lisbon, Portugal}
\affil[9b]{Laborat\'orio de Instrumenta\c{c}\~ao e F\'isica Experimental de Part\'iculas (LIP), Lisbon, Portugal}
\affil[10]{Institute for Astroparticle Physics, Karlsruhe Institute of Technology (KIT), Karlsruhe, Germany}
\affil[11]{Astrophysical Institute, Vrije Universiteit Brussel, Brussels, Belgium}
\affil[12]{Institute of Physics, Academia Sinica, Taipei, Taiwan}
\affil[13]{Department of Physics and Astronomy, University of California, Irvine, CA, USA}
\affil[14]{II Institute for Theoretical Physics, Hamburg University, Hamburg, Germany}
\affil[15]{Faculty of Mathematics and Natural Sciences, University of Wuppertal, D-42119 Wuppertal, Germany}
\affil[16]{INFN, Sezione di Firenze, Florence, Italy}
\affil[17]{School of Physics and Astronomy, Monash University, Clayton, VIC 3800, Australia}
\affil[18]{IPHC – Institut Pluridisciplinaire Hubert Curien, CNRS-IN2P3 / Université de Strasbourg, UMR 7178, Strasbourg, France}
\affil[19]{Institute for Space-Earth Environmental Research (ISEE), Nagoya University, Nagoya, Japan}
\affil[20]{Department of Physics, University of Jyväskylä, Jyväskylä, Finland}
\affil[20b]{Helsinki Institute of Physics, University of Helsinki, Helsinki, Finland}
\affil[21]{Deutsches Elektronen-Synchrotron (DESY), Platanenallee 6, 15738 Zeuthen, Germany}
\affil[22]{Erlangen Center for Astroparticle Physics (ECAP), Friedrich-Alexander-Universität Erlangen-Nürnberg, Nikolaus-Fiebiger-Straße 2, 91058 Erlangen, Germany}
\affil[23]{Department of Physics and Astronomy, University of Utah, Salt Lake City, UT 84112, USA}
\affil[24]{European Organization for Nuclear Research (CERN), Geneva, Switzerland}
\affil[25]{Instituto Galego de Física de Altas Enerxías (IGFAE), Universidade de Santiago de Compostela, Santiago de Compostela, Spain}
\affil[26]{Department of Physics, University of Arizona, Tucson, AZ, USA}
\affil[27]{Max-Planck-Institut für Kernphysik, Heidelberg, Germany}
\affil[28]{Department of Physics, Lund University, Lund, Sweden}
\affil[29]{SUBATECH – Laboratory of Subatomic Physics and Associated Technologies, University of Nantes, IMT Atlantique, CNRS/IN2P3, Nantes, France}
\affil[30]{Institute for Experimental Physics 1b, RWTH Aachen University, Aachen, Germany}
\date{\today}

\begin{document}
\pagenumbering{roman}
\maketitle

\begin{abstract}
    In high-energy and astroparticle physics, event generators play an essential role, even in the simplest data analyses. As analysis techniques become more sophisticated, e.g. based on deep neural networks, their correct description of the observed event characteristics becomes even more important. Physical processes occurring in hadronic collisions are simulated within a Monte Carlo framework. A major challenge is the modeling of hadron dynamics at low momentum transfer, which includes the initial and final phases of every hadronic collision. QCD-inspired phenomenological models used for these phases cannot guarantee completeness or correctness over the full phase space. These models usually include parameters which must be tuned to suitable experimental data. Until now, event generators have been developed and tuned mainly on the basis of data from high-energy physics experiments at accelerators. The wealth of data available from the latest generation of astroparticle experiments has not yet been fully exploited, and in many cases is not satisfactorily described. Both kinds of data sets are complementary as astroparticle experiments provide sensitivity especially to hadrons produced nearly parallel to the collision axis and cover center-of-mass energies up to several hundred TeV, well beyond those reached at colliders so far. In this report, we provide an overview of state-of-the-art event generators and their tuning, including the most relevant inputs from high-energy accelerator and astroparticle experiments. We present a road map that shows, for the first time, how the unified tuning of event generators with accelerator-based and astroparticle data can be performed.
\end{abstract}

\newpage
\pagenumbering{arabic}
\section{Introduction}

The simulation of high-energy particle collisions is an essential task in many fields of science, such as high-energy nuclear and particle physics or high-energy astroparticle physics. The simulation involves several steps, namely the event generation, hadronization and particle decay, stable particle propagation, and detector response simulation. The predictions of the event generators are usually based on the Standard Model (SM) of particle physics with new phenomena being tested against the SM predictions. So far, event generators are essentially developed and tuned solely based on data from accelerator-based experiments. \emph{Tuning} is the process of adjusting free parameters of phenomenological models in event generators based on comparisons with data. It differs from "fitting" in that the function being fitted is not arbitrary, but is given by the model. In this article, we explore the opportunities and challenges of incorporating data from astroparticle experiments into the development and tuning of event generators, and outline how such a \emph{global tuning} can be achieved. Given that specific tunes of event generators based on accelerator experiments have been found to be inconsistent with data from astroparticle experiments, there is a clear need for such an effort with potential benefits for a wide range of applications.

Event generators are crucial for many purposes, including the design of new experiments, the development of data analysis methods, or the prediction of particle interactions with detector material.
In addition, when searching for rare events or processes, such as Higgs production at the Large Hadron collider, data are typically contaminated by background, and event generators help to find experimental designs and analysis methods that allow one to reduce this contamination. For this purpose, they need to predict the frequency and distribution of signal and background events with a high degree of accuracy, since the efficiency and purity of an event selection can often only be calculated on the basis of a full end-to-end simulation of the entire experiment. The need for accurate event generators becomes more important as better experimental data become available, and is particularly important also for applying machine learning methods, which tend to outperform classical analysis, e.g. in classification performance, at the cost of being more sensitive to mismatches between simulated and experimental data.

In astroparticle physics experiments, event generators are used to simulate interactions of cosmic particles with the Earth and its atmosphere and in specific applications also to simulate particle interactions within cosmic ray sources and during their journey to Earth. High-energy cosmic rays, cosmic neutrinos, and gamma-rays are characterized by low fluxes, so that large aperture ground-based experiments are required to detect them. These experiments observe cosmic particles indirectly through showers of secondary particles. The showers themselves originate from collisions of high-energy primary particles in matter, most importantly air, water, or ice, and are typically detected by their emission of light (fluorescence or Cherenkov), radio emission, or classically by sparse arrays of charged particle detectors distributed over an extended area or volume. More recently, experiments have begun to combine two or more of the above methods to observe the same shower in different ways. This is known as \textit{hybrid detection}. Event generators simulate the development of the particle cascade to determine the relationship between the detector response and the initial cosmic particle. If the simulation is not accurate, the interpretation of data is biased. While the energy scale of TeV gamma-ray experiments can in principle be calibrated against the GeV gamma rays detected by satellite experiments using standard candles such as the Crab Nebula \cite{Bastieri:2005qi,Meyer:2010tta}, there is no equivalent calibration source for high-energy cosmic rays and neutrinos, making the theoretical uncertainties in the event generators a major source of uncertainty in these experiments. A prime example of this is the \textit{muon puzzle}~\cite{Albrecht:2021cxw} in extensive air showers (EAS), which causes an ambiguity in the inferred mass composition of ultra-high energy cosmic rays \cite{Kampert:2012mx}.

A major source of uncertainty in event generators is the treatment of hadronic interactions at low momentum transfer, where perturbative quantum chromodynamics (pQCD) is not applicable. The most important example here is the copious production of light-flavor particles which dominate the development of secondary particle cascades in various media. But also in high-momentum transfer interactions, such as  heavy-flavor production, the initial (parton momentum distribution) and final stages (hadronisation) are non-perturbative.

QCD-inspired phenomenological models are used to describe non-perturbative processes, which are neither guaranteed to be correct nor complete over the entire phase-space. The discovery of collective flow and enhanced strangeness production not only in nucleus-nucleus but also proton-proton and proton-lead collisions at the TeV scale \cite{Kalaydzhyan:2015xba,ALICE:2016fzo} are recent examples of surprises in QCD.

\begin{table}
    \centering
    \tablecaption{Key characteristics of HEP accelerator-based experiments and astroparticle experiments. We use $A$ as a placeholder for a nucleus, $\sqrt{s}$ is the energy in the nucleon-nucleon center-of-mass system.}
    \label{tab:comparison accelerator and astroparticle experiments}

    \begin{tabular}{lcccc}
        \toprule
                                      & \multicolumn{2}{c}{Accelerator}                                 & \multicolumn{2}{c}{Astroparticle}                                                                                \\
                                      & Fixed-target                                                    & Collider                                         & Cosmic rays                                 & Neutrinos       \\
        \midrule
        Collision energy ($\sqrt{s}$) & up to 100\,GeV                                                  & 100\,GeV to 14\,TeV                              & up to 500\,TeV                              & up to 10\,TeV   \\
        Collision systems             & \multicolumn{2}{c}{e+e, e+p, p+p, p+$A$, $A$+$A$} & \multicolumn{2}{c}{($\pi$,K,p,$A$,e,$\gamma,\mu,\nu$)+$A$}                                                                 \\
        Initial state                 & \multicolumn{2}{c}{fixed}                                       & \multicolumn{2}{c}{variable (energy and system)}                                                                 \\
        Final state: rapidity         & full                                                            & mid to forward                                   & \multicolumn{2}{c}{very forward}                              \\
        Final state: flavour          & light                                                           & light and heavy                                  & light                                       & light and heavy \\
        Resolution                    & \multicolumn{2}{c}{single interaction}                          & cascade                                          & \parbox{2.8cm}{single interaction, cascade}                   \\
        \bottomrule
    \end{tabular}
\end{table}

To arrive at a highly complete and accurate description of QCD phenomena, and to reduce uncertainties and ambiguities in the interpretation of data from astroparticle experiments, it is important to develop event generators using data from all sources. So far, data from astroparticle experiments have not been widely used in this context, apart from a few pioneering studies, see for example~\cite{Baur:2015gzu}. While previous generations of astroparticle experiments were not precise enough for this purpose, the latest generation provides a great wealth of data.

A comparison of specific aspects of accelerator and astroparticle physics data is given in \autoref{tab:comparison accelerator and astroparticle experiments}. Several challenges must be overcome to use astroparticle data: the initial state of a collision is variable and not well known, and air shower detector arrays observe the final state of a cascade of interactions in a medium such as air rather than a single interaction, while optical and radio observations observe the evolution of the entire shower particle ensemble along the shower axis. 
However, astroparticle data are complementary to accelerator data: they are sensitive to light and heavy flavor production at forward rapidities and probe collisions at center-of-mass (CM) energies of up to $500$\,TeV, as well as to collisions that are not easily accessible at colliders~\cite{Petrov:2009wr,Ryutin:2011kt}, such as those initiated by pions and kaons. By exploiting complementary data from both fields, theoretical uncertainties in event generators can be further reduced. This also provides a powerful test of the effective phenomenological models employed; ultimately, event generators including sufficiently general models should be able to describe all the data without inconsistencies.

Automatic tuning allows a rapid cycle of testing new models, and allows a quick retuning as new models or data become available. Due to the fundamental differences between the measurements in the two fields, this is a complex effort, mostly because of the unknown initial state of the first and subsequent collisions and the computational resources needed for full EAS simulations.
The development of standardized tools is a crucial prerequisite for this effort.  

In this report, we present for the first time a road map showing how to tune event generators simultaneously using data from both accelerators and astroparticle physics experiments. In \autoref{sec:event generators}, we summarize the theoretical approaches implemented in current event generators and recent developments towards event generators that are applicable to both high-energy and astroparticle physics. In \autoref{sec:particle transport}, we summarize recent developments in the simulation of particle transport in matter, which are essential for the interpretation of high-energy astroparticle experiments and for global tuning. In \autoref{sec:input from experiments}, we give an overview of the most important measurements from accelerator and astroparticle experiments that provide input for global tuning. In \autoref{sec:tuning}, we summarize the current state of tuning of event generators and discuss how the tools involved need to be extended or replaced to enable global tuning, followed by a discussion of current challenges and possible solutions to achieve global tuning in \autoref{sec:combination}. We conclude with a summary in \autoref{sec:summary}.

\section{Theoretical foundations and event generators}\label{sec:event generators}

\begin{table}[tb]
    \tablecaption{Comparison of five event generators. Acronyms are defined in the text.}
    \label{tab:generators}
    \centering
    \footnotesize
    \setlength{\tabcolsep}{3pt}
    \begin{tabular}{p{2.6cm}p{2.3cm}p{2.3cm}p{2.3cm}p{2.3cm}p{2.1cm}}\toprule
        & \epos{}4 \cite{Werner:2023zvo}  & \eposlhc{}-R \cite{Pierog:2023ahq}  & \qgsjet{}III \cite{Ostapchenko:2024myl}  & \sibyll{}~2.3d \cite{Riehn:2019jet}  & \pythia~8 \cite{Bierlich:2022pfr,Sjostrand:2021dal}
        \\ \midrule
        Primary domains                                                   & HIC, HEP                                                  & EAS, HIC                                 & EAS                             & EAS                             & HEP
        \\
        Theoretical basis                                         & parton-based GRT, pQCD,\n energy sharing,\n saturation        & parton-based GRT, pQCD,\n energy sharing & GRT, pQCD \n (DGLAP+HT)         & GRT, pQCD\n (minijet)           & MPI, pQCD,\n ISR, FSR
        \\
        Nuclear collisions                                        & idem                                                          & idem                                     & idem                            & extended \n superposition       & Glauber via \n Angantyr
        \\
        Pomeron                                                   & semi-hard,\n dynamical\n saturation                           & semi-hard                                & semi-hard                       & soft+hard                       & soft+hard
        \\
        Parton\n distributions                                    & generated                                                     & custom (GRV\n for valence)               & Pomeron PDFs \n + DGLAP + HT    & GRV                             & various
        \\
        Diffractive \n
        dissociation \n
        (low mass)                                                & diffractive Pomeron                                           & diffractive Pomeron                      & Good-Walker (3-channel eikonal) & Good-Walker (2-channel eikonal) & longitudinal \n strings
        \\
        Diffractive \n
        dissociation \n
        (high mass)                                               & Pomeron \n exchange                                           & Pomeron \n exchange                      & cut-enhanced \n graphs          & Pomeron \n exchange             & MPI
        \\
        String \n fragmentation                                   & area law                                                      & area law                                 & early Lund type                 & Lund                            & Lund
        \\
        Forward-central \n
        correlation                                               & strong                                                        & strong                                   & strong                          & weak                            & strong
        \\
        Charm production                                          & pQCD                                                          & parameterised \n + intrinsic                                      & ---                             & parameterised \n + intrinsic    & pQCD
        \\
        Collective effects                                        & core-corona,\n hydrodynamical flow,\n hadronic\n rescattering & core-corona,\n parameterised flow,\n hadronic\n rescattering       & ---                             & ---                             & colour \n reconnection,\n rope fragm., \n string shoving, \n hadronic\n rescattering
        \\
        $\lim_{s\rightarrow \infty}(\sigma_{\pi p}/\sigma_{p p})$ & 1                                                             & 1                                        & 1                               & 1                               & 2/3                                                                                  \\
        Programming\newline language                              & C/C++, \n Fortran90, \n Fortran                               & Fortran                                  & Fortran                         & Fortran                         & C++
        \\
        \bottomrule
    \end{tabular}
\end{table}

Despite significant progress both in the predictions of perturbative Quantum Chromodynamics (QCD) and in measurements at the Large Hadron Collider (LHC), it is still not possible to calculate from first principles the bulk of particle production processes at high energies. Only processes with large momentum transfer, also known as hard interactions, are accessible to the perturbative methods. To obtain a complete description of hadron collisions in accelerator experiments, it is necessary to combine results from perturbative QCD and general theoretical constraints with phenomenological modeling. To make predictions for particle production in hadron collisions in the astroparticle context, one must also extrapolate the distributions measured at accelerators into unmeasured regions of phase-space and to much higher energies.

Processes in which heavy quarks are produced in the final state are necessarily hard, and pQCD calculations can be quite accurate, but even in these calculations non-perturbative parts enter to model long-distance physics effects governing the parton-to-hadron transition and vice versa (fragmentation functions/hadronization, parton distribution functions).

Event generators use Monte Carlo simulations to describe QCD interactions. \autoref{tab:generators} presents the latest generation of event generators that are used in both high-energy accelerator physics and astroparticle physics. A brief description of each of these generators can be found in Appendix~\ref{app:had-models}.

\pythia~8~\cite{Bierlich:2022pfr} is an event generator often used in high-energy physics (HEP) and more recently also applied for cosmic ray physics~\cite{Sjostrand:2021dal}. Its core is built on the pQCD model. A (semi-)hard scattering is the starting point of each inelastic interaction, accompanied by emissions of partons before, or after the hard scattering (so-called initial/final state radiation implemented in parton shower algorithms). Multiple (semi-)hard scatterings are also allowed. The soft physics effects are described by phenomenological models. The cross sections for (semi-)hard scattering are determined from pQCD using a threshold in the transverse momentum (suppressed by a smooth damping factor). The modeling of the total elastic, diffractive, and inelastic cross sections in \pythia~8 is decoupled from the particle production mechanism.

This is different for the event generators used in astroparticle physics. Most of these describe, with varying degrees of rigor, the interaction of hadrons as the exchange of specifically structured "networks" of interacting quarks and gluons (so-called Pomeron and Reggeon exchanges). The mathematical framework for these exchanges, an effective quantum~field~theory, is the so-called Gribov-Regge theory~(GRT)~\cite{Gribov:1967vfb,Gribov:1968jf}, see Refs.~\cite{Levin:1997mg,Drescher:2000ha} for a pedagogical introduction. It allows one to connect the wide range of processes that occur in hadron collisions, such as elastic scattering, diffractive scattering and soft multi-particle production up to multiple hard parton scattering. The energy evolution of the hadron multiplicity and the total cross section are thus linked in GRT-based models.

Sibyll~\cite{Ahn:2009wx,Fedynitch:2018cbl,Riehn:2019jet} is a GRT model that describes inelastic collisions with a soft and a hard part, where the hard part is based on the pQCD cross section calculated with an energy-dependent $\pTo$ cutoff (similar to \pythia), while the soft part is purely phenomenological. The \epos{}~\cite{Werner:2023zvo,Pierog:2023ahq} and \qgsjet{}~\cite{Ostapchenko:2024myl} families of models use a semi-hard pomeron that consistently mixes aspects of the GRT and pQCD descriptions. The semi-hard pomeron provides the analog of initial and final state radiation (ISR and FSR), which in particular leads to broader hadron spectra in rapidity. The \epos{} family of generators is based on so-called parton-based GRT, where the pomerons are exchanged between partons instead of hadrons and energy conservation is ensured at the amplitude level. The \qgsjet{} family of generators implements pomeron-pomeron interactions. Both are mechanisms that slow down the growth of the total cross section at high energies, which is necessary to describe the measurements. 

Single diffraction dissociation, which occurs in events where only one of the incoming hadrons is dissociated, is an important contribution to hadron production in the forward region. In GRT models, this is described by the exchange of a specific colorless (quantum number-less) configuration of quarks and gluons (the so-called diffractive pomeron). This is either explicitly included in the amplitude or added using the Good-Walker model~\cite{Good:1960ba}.

In the context of air showers, the models need to be reliably extrapolated from hadron-hadron to hadron-nucleus and nucleus-nucleus interactions. In the \epos{} and \qgsjet{} models the nuclear amplitudes are constructed from the basic pomeron amplitude using the Glauber formalism~\cite{Glauber:1955qq,Glauber:1970jm}. In Sibyll, a mixture of Glauber (for hadron-nucleus) and an extended superposition model (for nucleus-nucleus)~\cite{Engel:1992vf} is used. \pythia can use the Angantyr module \cite{Bierlich:2018xfw}, which superimposes individual nucleon-nucleon interactions following a Monte-Carlo Glauber calculation. \pythia can optionally use nuclear parton distribution functions (PDFs) for collisions with nuclei, which differ from the PDFs of free protons.

The description of heavy ion collisions (HIC) requires the inclusion of effects such as collective flow, jet quenching, and possible modifications of the final-state hadron composition (which may affect e.g.\ the strangeness enhancement). In \epos{}, this is described by distinguishing between hadronization of the high-(energy) density part of the collision (core), modeled by the formation of a quark-gluon plasma that evolves hydrodynamically and then decays statistically, and hadronization of the low-(energy) density part (corona), based on string fragmentation (similar to the standard \pythia hadronization mechanism). This is followed by a phase of hadronic rescattering. Similarly, in \pythia~8, color reconnection, rope fragmentation, string shoving, and hadron rescattering are means to modify standard hadronization in vacuum to describe the effects observed in HIC. \sibyll{} and \qgsjet{} do not include these types of HIC effects.

Heavy quark production has traditionally not been implemented in event generators used for EAS simulation, since the effect on most air shower observables is negligible~\cite{1995ICRC....1..123K}, but the need to model the prompt atmospheric lepton flux in the latest generation of neutrino observatories has led to the inclusion of charm production in \sibyll{}-2.3d, \epos{}4, and \eposlhc{}-R. \pythia can produce all heavy quarks.

Event generators are stand-alone codes often with non-standard interfaces. However, software packages such as CRMC~\cite{2021zndo...5270381U} and Chromo~\cite{Dembinski:2023esa} simplify the use and comparison of the previously listed event generators by providing a common interface and unified output.

\section{Particle transport in matter}\label{sec:particle transport}

\begin{table}
    \tablecaption{Comparison of transport codes in the context of tuning. The CPU time estimates are order-of-magnitude and valid only for a fast hadronic interaction model like \sibyll, and without simulation of Cherenkov or radio emission.}\label{tab:transport codes}
    \small
    \begin{tabular}{p{3.5cm} p{3.4cm} p{3.4cm} p{3.4cm}}
        \toprule
                                                              & Monte-Carlo \newline simulation                           & Cascade equation \newline solver    & 3D Hybrid simulation  \\ \midrule
        Examples                                              & \corsika{}, \crpropa{}                                          & \conex{}, \mceq{}                         & \corsika{}+\conex{}, \seneca{}                       \\
        Observables                                           & all                                                       & limited (see text)                  & all                                         \\
        Shower-to-shower \newline fluctuations?               & yes                                                       & no                                  & yes                                         \\
        Cost of calculating observables for 10 EeV air shower & 10k CPU hours for \newline 1000 air showers with thinning & (1-100)k CPU hours for all energies (pre-computing tables) & 2k CPU hours \newline for 1000 air showers  \\
        Cost of calculating observables for 10 PeV air shower & 1000 CPU hours \newline for 1000 air showers  & (1-100)k CPU hours for all energies (pre-computing tables) & 200 CPU hours \newline for 1000 air showers \\
        Energy dependence of computational cost               & $\propto E_\text{cosmic ray}$                             & $\approx$ constant                  & $\propto E_\text{cosmic ray}$   \\
        \bottomrule
    \end{tabular}
\end{table}

Transport codes simulate the propagation, decay, and interaction of high-energy particles with a medium such as air, water, ice, or interstellar gas. They employ event generators to handle interactions and decays. This section provides an overview of the most relevant particle transport codes and some of their applications. A recent review of transport codes is also given in Ref.\,\cite{Albrecht:2021cxw}. In \autoref{tab:transport codes}, transport codes are compared from the point of view of their use in event generator tuning. They link the physics of hadronic interactions in a cascade with astrophysical \enquote{macroscopic} experimental observables, such as the depth of the shower maximum $X_\text{max}$, the number of muons produced, or the atmospheric high-energy lepton flux. Theoretical uncertainties in particle transport arise mainly from the theory and the phenomenological assumptions implemented in the event generators, but are also related to the propagation of secondary particles. The propagation of high-energy muons through dense media has recently been significantly improved with the \proposal{} propagation code \cite{Koehne:2013gpa,Dunsch:2018nsc}. Such effects are important in neutrino observatories like IceCube and in underground laboratories.

From the point of view of an accelerator-based experiment, transport codes (such as GEANT4~\cite{GEANT4:2002zbu} or \fluka{}~\cite{Bohlen:2014buj}) simulate particle transport through \enquote{detector material}. For example, the Earth's atmosphere and magnetic field can be considered as part of the detector of an air shower experiment, and must to be carefully monitored, as these properties change on a daily and seasonal basis. Tracking the vertical density profile of the atmosphere and its optical properties is important, which is done using laser and electron beams \cite{PierreAuger:2013dlv,TelescopeArray:2013jga}, local weather balloon flights and satellite-based models \cite{PierreAuger:2012jsu}. In accelerator-based experiments, theoretical uncertainties are reduced by minimizing the amount of material the particles must pass through and by sophisticated calibration schemes. In astroparticle experiments, the amount of material is necessarily large, and calibration against a known source is not possible, except at gamma-ray observatories. In this case the Crab nebula is often used as a reference source. Therefore, great care is taken to make not only the interaction but also the transport codes as accurate as possible. The impact of theoretical uncertainties in event generators is further amplified for some air-shower observables, such as the total number of produced muons, which depends on the evolution of the whole hadronic cascade, while other observables depend mainly on the first few interactions of the primary cosmic ray, such as the depth of the shower maximum.

Computational speed and efficiency of transport codes is a key aspect to be considered in  tuning procedures, since calculating a change in an air shower observable as response to a change in an event generator may require simulating hundreds of air showers in order to average out shower-to-shower fluctuations. Full Monte Carlo simulation, as employed by \corsika{}~\cite{Heck:1998vt} and \crpropa{}~\cite{AlvesBatista:2022vem}, is the gold standard, but very demanding in terms of computational resources. The computational effort is proportional to the number of particles that need to be tracked. For air showers, this number is proportional to the cosmic-ray energy, which in turn spans over 11 orders of magnitude. Full simulation becomes impractical above $10^{17}$\,eV. At higher energies, the thinning technique \cite{Hillas:1997tf} in which only a representative subset of particles is simulated can be used. The computational load further increases if detailed Cherenkov or radio emission has to be simulated, which is needed for some observables.

The numerical solution of cascade equations is an alternative technique for simulating air showers and calculating particle fluxes. The latter is used  by \conex{} and \mceq{}. In this approach, hadronic cascades are described by a large system of differential (cascade) equations, one for each particle species, containing source and sink terms describing energy loss, particle interaction and decay
\cite{Gaisser:2016uoy}.
The input particles are binned in energy and atmospheric depth, and the differential equations are then solved numerically. This allows one to compute atmospheric lepton fluxes, for example, using as input the cosmic ray flux at the top of the atmosphere, or an \enquote{average} air shower from monoenergetic cosmic rays. Examples of outputs are longitudinal density profiles and energy distributions of secondary particle fluxes. This approach works for calculating only some of the air shower observables. Observables that require simulation of shower-to-shower fluctuations or lateral particle distributions at the detector sites (typically required by ground-based air shower arrays), require full Monte Carlo or 3D hybrid simulations, in which the initial and final stages of an air shower are Monte Carlo simulated, and cascade equations are solved numerically for the middle part \cite{Ortiz:2004gb,Pierog:2011kzx}. On the other hand, observables that are already defined as an average over many similar showers (such as the depth of the shower maximum or the number of muons produced) are obtained directly from these cascade equations solvers.

Numerical solvers of cascade equations use precomputed tables of the energy spectra of secondary particles for colliding particles at different energies. With these tables, it takes only seconds to compute an air shower, but the computational cost of generating these tables from event generator calls is substantial, requiring on the order of (10--1000)k events, depending on the desired smoothness of the tables and whether charmed hadrons, which have a lower production cross section, are to be simulated. However, cascade equation solvers have a computational advantage in tuning procedures where air shower observables of different primary cosmic ray energies and masses are to be computed multiple times during the tuning process.

The transport codes shown in \autoref{tab:transport codes} are briefly discussed in Appendix~\ref{app:transport} in the context of event generator tuning. 

Factorization or reweighting techniques can be used to reduce the computational cost of CPU-intensive transport codes. In the factorization approach \cite{Ulrich:2010rg,Baur:2019cpv}, intermediate \enquote{high-level} key variables are identified, such as the inelastic cross section or the hadron multiplicity, which have a strong influence on the air shower observables. Assuming that the high-level quantities scale logarithmically with the cosmic-ray energy, one can pre-calculate the impact on the air-shower observables.
This drastically reduces the effort to the calculation of a high-level variable by an event generator at a fixed cosmic-ray energy. An example of such an analysis is discussed below (\textit{cf.} Fig.\,\ref{fig:impact_study}).
On the other hand, in the reweighting approach, weights are applied to the precomputed tables used by a cascade equation solver, or to a set of previously Monte Carlo simulated air showers, which must be stored as complete graphs, including all particle interaction and decay history. The weights are chosen to reflect the change in the event generator. Reweighting can be effective when the change in the event generator has only an isolated effect, for example, only on selected particle types. Both strategies have the potential to speed up tuning considerably, but at the cost of making strong assumptions that need to be carefully validated. They also run the risk of missing unexpected side effects of changes to event generators.

\section{Input from experiments}
\label{sec:input from experiments}

Traditionally, event generators are only tuned to input from accelerator-based experiments (classic tuning), and only a few exceptional studies already explored more global tuning~\cite{Baur:2015gzu}. Classic tuning uses measurements of, e.g., production cross sections, hadron multiplicity spectra, energy flow, and rapidity gaps. For a global tuning a variety of air shower observables are available: the mean and fluctuations of the depth of the shower maximum, the mean and fluctuations of the produced number of muons, and other observables related to these. The mean number of muons is sensitive to small changes in secondary hadron yields, which become exponentiated by the hadronic cascade \cite{Ulrich:2010rg,Albrecht:2021cxw}, while the other air shower observables are dominated by the first interaction. Further input can be derived from the atmospheric muon flux at PeV energies, whose conventional component (i.e.\ the component arising from light-flavour production, hadronization, and decay) is sensitive to hadron production over a wide range of energies, while the prompt component (i.e.\ that arising from heavy-flavour production) is sensitive to charm production. Tuning to air shower observables also requires a model of the cosmic ray flux and its composition, which has its own uncertainties. These uncertainties in turn largely derive from uncertainties in the event generators, so a truly global tune should also adjust the cosmic ray flux model.

Inputs from air showers and accelerators complement each other, and both have their respective limitations. The highest CM energy at an accelerator on Earth achieved so far amounts to 13.6 TeV. In air showers, the CM energy of the first collision can be significantly higher, up to several hundred TeV. In many cases, the precision of air shower measurements is significant and puts constraints on quantities only loosely constrained by accelerator measurements. For example, the fluctuations in the muon number in air showers constrain the elasticity, the fraction of energy taken by secondary particles, which is difficult to measure at the LHC.

Another complementarity is in the accessible rapidity range. Measurements from accelerator-based experiments are often expressed as a function of the transverse momentum $\pT$, and rapidity $y$ or pseudorapidity $\eta$ of the products in the CM frame. Rapidity $y=\frac{1}{2}\ln[(E+p_\parallel) / (E-p_\parallel)]$, where $p_\parallel$ is the momentum component parallel to the beam line, is a measure of how forward going the particle is.

At the TeV scale, the production cross sections for hadrons near projectile rapidities are most important for air showers \cite{Albrecht:2021cxw}, but the most detailed measurements with identified hadrons from LHCb \cite{LHCb:2014set} end at $y < 5$. Further forward, we have measurements from TOTEM on the charged particle multiplicity \cite{CMS:2014kix}, and from CMS-CASTOR on the energy flow~\cite{CMS:2019kap}, and at $y > 8$ from the LHCf experiment~\cite{LHCf:2008lfy} on the $\piz$ and neutron production cross sections. There are no collider data on strangeness or charm production beyond $y > 5$. Muon production in air showers is sensitive to these production cross sections, and therefore can constrain the cross sections in the forward region.

Finally, there is complementarity in the collision systems. At the LHC, so far only \ppColl, \pPbColl, and \PbPbColl collision systems were investigated, while the most common system in air showers is $\pi$+(N,O). The properties of \piAColl collisions beyond the CM energies of fixed target experiments have to be inferred from data on \pAColl collisions, since pion beams are not available at colliders. Until measurements with the oxygen beams become available at the LHC, the extrapolation from \ppColl and \pPbColl to \pOColl and \pNColl remains uncertain at the TeV scale.

Global tuning will inevitably reveal discrepancies between models and data. On the experimental side, there can be hidden systematic effects in the measurements, not covered by the quoted uncertainties. On the theory side, models may lack the necessary physics content, robustness and flexibility to reproduce all available measurements. In either case, a good fit in the statistical sense of the models to the available data will not be possible, but global tuning is necessary to reveal these discrepancies, so they can be addressed by the community.

\subsection{Accelerator-based experiments}\label{sec:input from accelerator experiments}

Measurements at particle accelerators tackle properties of single interactions in different combinations of particles and nuclei and thus provide valuable constraints on the modeling of  fundamental interactions in the evolution of air showers. Accelerator-based experiments allow us to measure billions of collisions with a well-defined identical initial state and a fixed collision energy. This allows one to compute production cross sections to high accuracy and to measure relatively rare processes. On the flip-side, there are only a few beam-beam combinations at discrete energies available, so it is not possible to directly measure all relevant interactions which happen in air showers. Theory embedded in event generators will always be needed to inter- and extrapolate from these reference measurements.

In the following, we will mostly focus our discussion on experiments using hadron beams such as the ones provided at the LHC and its pre-accelerator the Super Proton Synchrotron (SPS). Nevertheless measurements at the Brookhaven Relativistic Heavy Ion Collider (RHIC) as well as older data from the Tevatron and other accelerators also provide important input by constraining event generators on a broad range of collision energies. Note that experiments conducted at lepton-lepton and lepton-hadron colliders such as LEP and HERA are also included in the tuning because they allow fixing string fragmentation parameters \cite{Pierog:2013ria} and the parton distribution functions of hadrons \cite{Bailey:2020ooq}, respectively. Future machines such as the Electron Ion Collider~\cite{eic} and the FCC-ee~\cite{fcc-ee} at CERN will probe hadrons at new scales. 

Accelerator data are available from either fixed-target or colliding-beam experiments. The major differences between these groups of experiments are in the CM energy scale that is being probed, the kinematic coverage, and the flexibility in studying different collision systems. Current fixed-target experiments probe nucleon-nucleon CM energies up to about 100\,GeV, the LHC experiments in colliding-beam configuration currently reach up to 13.6\,TeV.

An advantage of fixed-target experiments is the possibility to study a greater variety of beam and target combinations. The target is usually a thin foil or small block of material that can be easily changed. Liquid and gaseous targets are also used. Experiments can use primary beams, like the proton and lead beams from the LHC or SPS, or secondary beams produced by colliding the SPS beam with a production target. This is currently the only way to study \piAColl interactions.

Fixed-target experiments measure the produced particles in the rest frame of the target, \ie in the laboratory system. Some fixed-target experiments, like NA61/SHINE \cite{Abgrall:2014fa} ($|y| \lesssim 3$), cover most of the rapidity range in the CM frame, since the rapidity range is fairly narrow at these energies. Collider experiments typically measure particles in a frame that is close or identical to the CM frame of the colliding nucleons. Such experiments are typically most densely instrumented in the mid-rapidity region, where most particles are produced. On the other hand, the forward region ($y \gtrsim 2$), where most of the beam energy flows into, is typically sparsely instrumented.

The four large LHC experiments are ATLAS~\cite{ATLAS:2008xda}, CMS~\cite{CMS:2008xjf}, ALICE~\cite{ALICE:2008ngc}, and LHCb~\cite{LHCb:2008vvz}. They are all designed as general-purpose detectors, but each with a different focus. ATLAS and CMS have been designed for discovering new heavy particles, such as the Higgs boson and heavy supersymmetric candidates. This is best done in the mid-rapidity region ($|y| < 2.5$), where these experiments have their main instrumentation. They have lepton-hadron separation capabilities, but only limited hadron-identification capabilities. The main focus of ALICE is studying QCD in the mid-rapidity region, especially in proton-nucleus and nucleus-nucleus collisions. ALICE is well-equipped for this task with a high-resolution tracker and excellent hadron identification capabilities over a large momentum range. The main focus of LHCb is the study of the production of hadrons containing heavy flavor, i.e. charm and bottom quarks. It is instrumented mainly in the forward region and has very good particle identification capabilities for hadrons with momenta up to 100\,GeV$/c$. 
LHCb is equipped with a system that enables the injection of gases into the beam pipe, allowing LHCb to operate as a fixed-target experiment~\cite{LHCb-PAPER-2014-047}. This provides a unique opportunity to study proton-nucleus and nucleus-nucleus collisions with different gaseous targets using the LHC beams. The system has recently been upgraded to extend the injected gases beyond noble gases, most notably also including nitrogen and oxygen~\cite{CERN-LHCC-2019-005}. 

Important observables in these experiments are total and differential cross sections, multiplicity distributions, and energy flow. Results from proton-proton, proton-nucleus and nucleus-nucleus collisions allow one to understand nuclear and collective effects. The relevance of the different quark flavors can be probed by studying the production of particles containing strange, charm or bottom quarks. A comprehensive overview of data from accelerator experiments that is useful for tuning hadronic interaction models is given in Ref.~\cite{Albrecht:2021cxw}. 
The most important experiments for the tuning of QCD-inspired models and their most relevant results are summarized in Appendix~\ref{app:exp}. 

\vspace{1mm}
For air shower simulations, it is important to understand nuclear effects, which modify production cross sections relative to collisions of free nucleons. So far, LHC beams have probed proton-proton, proton-lead, and lead-lead collisions, which represent extreme ends in the space of collision systems. In addition to that, a short pilot run with xenon beams was conducted in 2017. These reference systems are not ideal for air showers, in which collisions with nitrogen and oxygen are dominant. These nuclei have masses near the geometric mean between proton and lead, and it is not clear how features of hadron production shall be interpolated to this point. Present event generators used to simulate air showers show a considerable spread in their predictions for the hadron multiplicity in proton-oxygen collisions, at the level of $25\%$. In contrast, they agree to $5\%$ in proton-proton collisions in the mid-rapidity region, because they are tuned to these data \cite{Brewer:2021kiv}. 

Proton-oxygen and oxygen-oxygen collisions will allow us to determine how hadron production evolves from light to heavy systems at high energies, and are direct references for common air shower collisions. Comparing them will also allow us to test the superposition model assumption at the TeV scale. First runs with the LHC providing p+O and O+O collisions are planned for 2025~\cite{Slupecki:2888741}. There is a widespread interest in accelerating other nuclei at the LHC, and work towards this goal is coordinated by the Working Group on Future Ions at CERN~\cite{lilhcw2024}.

\subsection{Astroparticle experiments}\label{sec:input from astroparticle experiments}

Astroparticle experiments measure high-energy gamma rays, neutrinos, or cosmic rays, whose flux as a function of energy follows a steeply falling power law. Therefore, large apertures are needed for measurements at very high energies. This is achieved with ground-based experiments that measure the characteristics of particle showers initiated by the primary projectile in the Earth's atmosphere, water bodies or ice shields.

The properties of the primary projectile are inferred from the observed features of the particle shower. The direction of the projectile is inferred from the arrival times of shower particles at ground level detectors. The energy is derived from integration of the energy deposition profile in the atmosphere, which is a nearly model-independent technique, or from the number of particles at the ground level. The latter requires a detailed simulation of the particle cascade to obtain a calibration curve from the simulation.

\begin{figure}[tb]
    \centerline{\includegraphics[width=0.8\textwidth]{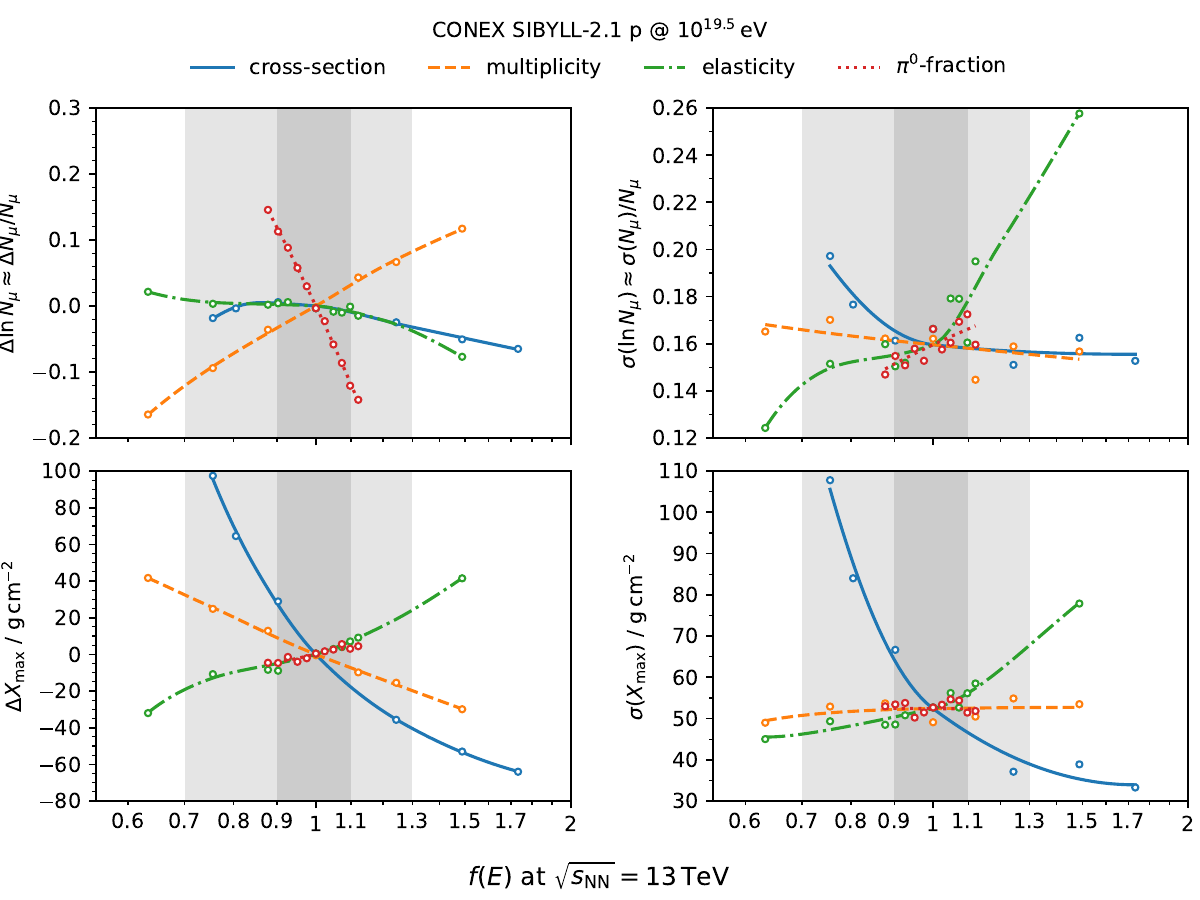}}
\caption{Effect of changing basic parameters of hadronic interactions on the means and standard deviations of the logarithm of the muon number \nmu (top row) and the depth \xmax of the shower maximum (bottom row) for a $10^{19.5}$\,eV proton shower simulated with \conex using \sibyll{}2.1 as the baseline model.  The left and right columns show relative shifts from the mean and fluctuations, respectively. The data, originally based on\cite{Ulrich:2010rg}, are shown as a function of the modification in the nucleon-nucleon system at a CM-energy $\sqrtsnn = 13$\,TeV, which is extrapolated logarithmically towards higher energies. The shaded bands highlight a $\pm 10\,\%$ and $\pm 30\,\%$ modification, respectively (Plot taken from \cite{Albrecht:2021cxw}).}
\label{fig:impact_study}
\end{figure}

The mass $A$ of a cosmic-ray projectile is inferred from the depth of the electromagnetic shower maximum $\xmax$, the depth of the maximum muon production $\xmumax$ \cite{Cazon:2012ti} or the number of muons $\nmu$ produced in the particle shower. Other mass sensitive observables can be reduced to these three fundamental observables. They probe different aspects of the hadronic interactions in the cascade, whose evolution is dominated by soft QCD interactions. Complementary information is obtained from the mean values of $\xmax$, $\nmu$, and $\xmumax$, and from their stochastic shower-to-shower fluctuations in a narrow energy interval. The fluctuations are very sensitive to the first interaction of the primary projectile, which is also true for $\avg\xmax$, while $\avg\nmu$ and $\avg\xmumax$ are sensitive to the whole evolution of the hadronic cascade. One can therefore probe QCD features of the first interaction at hundreds of TeV in the CM frame, and features of the whole chain down to GeV. \autoref{fig:impact_study}, adapted from \cite{Ulrich:2010rg,Albrecht:2021cxw}, illustrates in which way air shower observables, such as the muon number, \nmu, and the shower maximum position in the atmosphere, \xmax, react to changing basic parameters of hadronic interactions. For example, increasing the interaction cross section shifts \xmax to higher altitudes and reduces its fluctuations, but has only little effect on the muon numbers and their fluctuations. 
\autoref{tab:sensitivity of air shower observables} summarizes how the observables relate to QCD features and cosmic rays properties. In the case of the muon number, $\nmu$, further information can be obtained from the lateral distribution function, whose characteristic width is inversely proportional to the muon detection threshold.

\begin{table}[tb]
    \centering
    \tablecaption{Observables in astroparticle experiments (described in the text) and their sensitivity to QCD features and properties of the cosmic rays \cite{Ulrich:2010rg}. Legend: $\sinel$ inelastic $(p,\pi)$-air cross section, $\mult$ average number of secondary hadrons, $\inelast$ inelasticity (energy fraction carried by most energetic product), $\piefrac$ energy fraction carried by neutral pions, $\sigdb$ production cross section for D and B mesons, $C_1$ mean free path to first collision, $E$ primary energy, $\mlna$ mean logarithmic mass, $\slna$ standard deviation of logarithmic mass.}
    \label{tab:sensitivity of air shower observables}
    \begin{tabular}{l | p{0.8cm}p{0.8cm}p{0.8cm}p{0.8cm}p{0.8cm} | p{0.8cm}p{0.8cm}p{0.8cm}p{0.8cm}}
        \toprule
        Observable                                & \multicolumn{5}{c|}{Sensitivity to QCD feature} & \multicolumn{4}{c}{Sensitivity to cosmic ray feature}                                                                                            \\
                                                  & $\sinel$                                        & $\mult$                                               & $\inelast$ & $\piefrac$ & $\sigdb$   & $C_1$      & $E$        & $\mlna$    & $\slna$    \\
        \midrule
        $\int (\text{d}E/\text{d}x) \, \text{d}x$ &                                                 &                                                       &            &            &            &            & \checkmark &            &            \\
        $\avg\xmax$                               & \checkmark                                      & \checkmark                                            & \checkmark &            &            & \checkmark & \checkmark & \checkmark &            \\
        $\sigma(\xmax)$                           & \checkmark                                      &                                                       & \checkmark &            &            & \checkmark & \checkmark & \checkmark & \checkmark \\
        $\avg\xmumax$                             & \checkmark                                      & \checkmark                                            & \checkmark &            &            & \checkmark & \checkmark & \checkmark &            \\
        $\sigma(\xmumax)$                         & \checkmark                                      &                                                       & \checkmark &            &            & \checkmark & \checkmark & \checkmark & \checkmark \\
        $\avg\nmu$                                &                                                 & \checkmark                                            &            & \checkmark &            &            & \checkmark & \checkmark &            \\
        $\sigma(\nmu)/\nmu$                       &                                                 &                                                       & \checkmark & \checkmark &            & \checkmark & \checkmark & \checkmark & \checkmark \\
        \midrule
        $\Phi_{(\nu,\mu),\text{conv}}$            & \checkmark                                      & \checkmark                                            & \checkmark & \checkmark &            & \checkmark & \checkmark & \checkmark &            \\
        $\Phi_{(\nu,\mu),\text{prompt}}$          &                                                 &                                                       & \checkmark &            & \checkmark & \checkmark & \checkmark & \checkmark &            \\
        \bottomrule
    \end{tabular}
\end{table}

While the direction and energy of a cosmic ray can be determined in a model-independent way, other observables are sensitive to the \textit{a priori} unknown mean and variance of the logarithmic mass number $\lna$. This additional uncertainty poses a challenge to the tuning of event generators, but can be addressed in two ways. First, there is a unique energy window in the cosmic-ray flux of (1--3)\,EeV, where the primary mass is dominated by protons \cite{PierreAuger:2012egl}. The proton fraction can be further enhanced by selecting showers with a deep shower maximum, thus avoiding the uncertainty in the composition. Second, one can study several observables simultaneously, all of which must be consistent with the same $(\mlna, \slna)$ hypothesis. The muon deficit in air shower simulations was discovered in this way \cite{PierreAuger:2014ucz,PierreAuger:2021qsd}.

Neutrino telescopes, measuring the flux of atmospheric leptons and astrophysical neutrinos, provide additional observables for tuning. The atmospheric lepton fluxes are the result of the cascade generated by cosmic-ray interactions with the atmosphere, as discussed in \autoref{sec:event generators}.  At low energy, the conventional flux due to the decay of pions and kaons in the atmosphere dominates. At high energies, the prompt flux then takes over. It originates from the decay of short-lived hadrons containing charm and bottom quarks, generally produced in the first few interactions of the primary cosmic ray. In the case of the flux of muons other short-lived resonances also contribute. Although most charm mesons are produced at mid-rapidity, due to the steeply falling cosmic-ray flux, charm mesons produced at rapidities $> 4.5$ contribute most to the prompt neutrino flux. Thus, the atmospheric lepton flux is sensitive to QCD features over a wide range of energies. At the level of a single neutrino interaction event, one cannot distinguish between the atmospheric and the extraterrestrial neutrino fluxes, but the combined neutrino flux can still be used for tuning since both sources contribute differently at different energies. On the other hand, the atmospheric and extraterrestrial neutrino fluxes are expected to present different features and shape, and the correlations with sources could become a powerful tool to distinguish them. This way, the perspective is open that each of these two fluxes could be separately used for tuning purposes.  

In summary, there is a wealth of data from astroparticle experiments available for tuning. The experimental precision achieved in modern air-shower experiments is competitive with accelerator-based experiments and challenges event generators tuned to LHC measurements. To disentangle the complex dependencies between the microscopic properties of hadronic interactions and the \textit{macroscopic} observables in these experiments, several observables should be used together in the tuning, while carefully considering the systematic uncertainties and correlations in the measurements. Hybrid experiments such as the Pierre Auger Observatory and IceCube, which can measure several observables simultaneously, offer the greatest value for tuning. 
A detailed discussion of the most relevant astroparticle experiments and their recent measurements in the context of event generator tuning is given in Appendix~\ref{app:exp}. 

\begin{figure}[tb]
    \centering
    \includegraphics[width=0.75\textwidth]{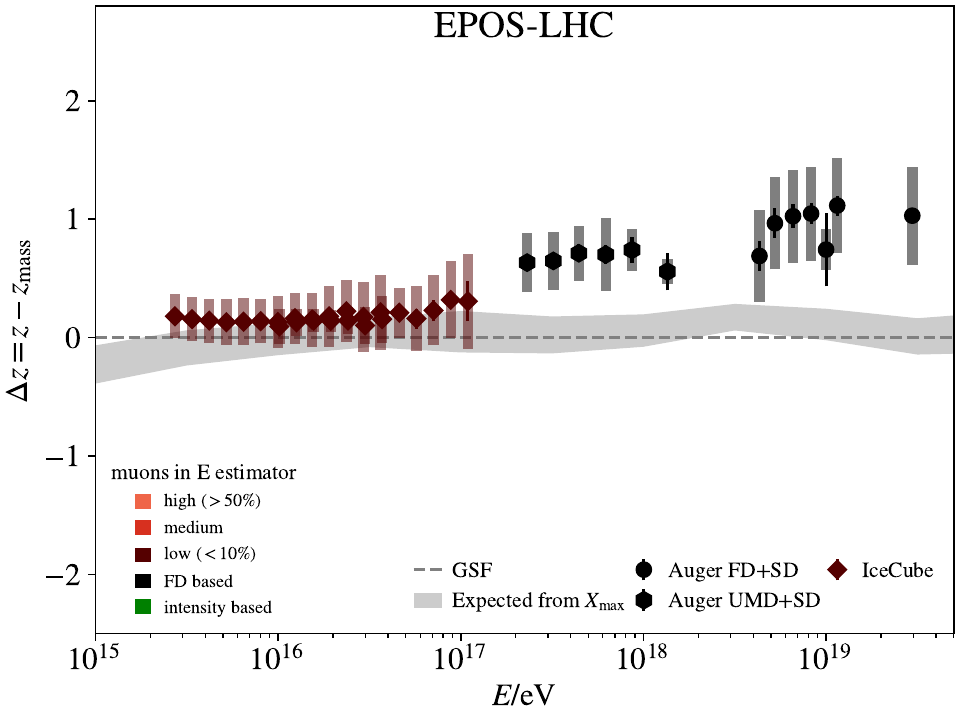}
    \caption{Muon content of air showers encoded in $\Delta z$-values (see text) as a function of the shower energy $E$ from different experiments. Only experiments with little (red-brown) or no (black) muon contribution to the energy estimator are shown. The $\Delta z$ value shows the deviation of the muon content from the expectation based on the data driven GSF model~\cite{Dembinski:2017zsh} and the event generator \eposlhc. The gray band indicates the expectation when the mass is inferred from Auger \xmax measurements instead of GSF. Error bars show statistical and systematic uncertainties added in quadrature. Figure adapted from Ref.~\cite{ArteagaVelazquez:2023fda}}.
    \label{fig:WHISP}
\end{figure}

\subsection{Clash of high-energy accelerator and astroparticle physics: the muon puzzle}
\label{sec:muon-puzzle}
A prominent example that illustrates the need for a coordinated effort by both the particle physics and the astroparticle physics communities is the so-called \textit{muon puzzle} in air shower data~\cite{Albrecht:2021cxw}. In the last decade an increasing number of datasets have revealed a consistent systematic discrepancy between the number of muons observed in EAS with respect to those predicted by standard interaction models. This gap persists despite data from the LHC having been included in the tuning of the hadronic interaction models. 

Apart from the cosmic ray energy $E$, the observed muon density at ground level depends on the atmospheric depth of the ground array, the lateral distance at which the muon density is measured, the zenith angle of the showers considered, and the effective energy cutoff introduced by the shielding of the detectors. Due to the diversity of the measurements the Working Group for Hadronic Interactions and Shower Physics (WHISP) consisting of representatives from most of the experiments was formed~\cite{EAS-MSU:2019kmv,Cazon:2020zhx,Soldin:2021wyv,ArteagaVelazquez:2023fda}.

To facilitate the comparison of observed vs.\ expected muon numbers for the very diverse experimental conditions, the group defined an abstract $\Delta z$-scale. The $\Delta z$-values are proportional to the difference between the observed number of muons in an experiment and the expectation calculated with an event generator and a model for the primary flux. An overview of all the muon measurements is presented in Appendix~\ref{app:exp-muon-puzzle}.

At first glance, a globally coherent picture from 1 PeV to 10 EeV does not emerge. However, two groups of experiments can be identified. Experiments that determine the shower energy largely or almost independently of the muon number, such as IceCube and Auger, show a muon deficit in the simulations that grows at a constant rate with increasing energy (see \autoref{fig:WHISP}). On the other hand, for experiments without independent energy measurements, no clear picture emerges (see Appendix~\ref{app:exp-muon-puzzle}). presumably because the shower energy is reconstructed (at least in part) from the number of muons, thereby masking the deficit.

In addition to the scientific interest in the nature of hadronic multiparticle production at high energies, the muon puzzle unfortunately introduces large systematic uncertainties in the analysis and interpretation of EAS data. This is particularly critical when training machine learning algorithms with simulations that are not fully consistent with the data. 

\section{Tuning of event generators}\label{sec:tuning}

Event generators use effective models with tens to hundreds of parameters that need to be adjusted to experimental data, in a process called \emph{tuning}. A \emph{tune} refers to one given set of these parameters. Many event generators have switches to enable or disable physics processes or select alternative physics models, which are also stored in the tune. Depending on the enabled processes, a tune can be more conservative or more speculative.

A prerequisite for tuning is the availability of experimental measurements available in a machine-readable format. In the particle physics community, this is provided by the High-Energy Physics Database (HEPData)~\cite{Maguire:2017ypu}, an open-access repository for sharing data from experimental particle physics. A project with a similar goal in the astroparticle community is the Cosmic Ray Database (CRDB)~\cite{Maurin:2023alp}. Then, one has to compare the output of an event generator to the measurement, in order to compute a residual. This step is not trivial due to the different internal implementations chosen by each event generator and the diversity of measurements.

Event generators suitable for tuning provide a documented interface that allows one to change parameters without recompiling the code. \pythia~8 is exemplary in this regard. Other event generators, including most of those used in the astroparticle physics community, are not designed to allow for tuning from a generic user. The parameters are hard-coded or the tuning interface is not documented. Tuning may be done manually by changing parameters until the generator predictions match the experimental results in the control plots. Alternatively, automatic tuning by performing a multi-dimensional fit makes results more reproducible and eases the incorporation of new data. 

\subsection{Automatic tuning}
\label{sec:autoTuning}
In automatic tuning, a suitable distance measure, such as a least-squares-type cost function, is used to quantify the agreement of the event generator output with a set of measurements, taking into account experimental uncertainties. The best set of parameters is found either by minimizing the least-squares-type cost function via gradient descent, or in a Bayesian framework by computing the posterior probability density of the parameters from the likelihood function and priors. Further details on these methods are presented in \autoref{app:tuning}.

Before tuning, the raw output of an event generator must be converted into a prediction that can be compared to the measurement. In particle physics this step is performed by the \rivet~\cite{Bierlich:2019rhm} software which has been specifically designed to support the development, validation and tuning of event generators. 
The experimental data are entered via so-called \rivet plugins which are programs that emulate the published experimental analysis but starting from the generator output, instead of the original experimental data. \rivet stores measurements in a human readable format, usually imported from HEPData. Some event generators, such as \pythia~8, can feed the raw events directly into \rivet, while others use interface programs~\cite{Dobbs:2001ck,2021zndo...5270381U,Dembinski:2023esa}. 

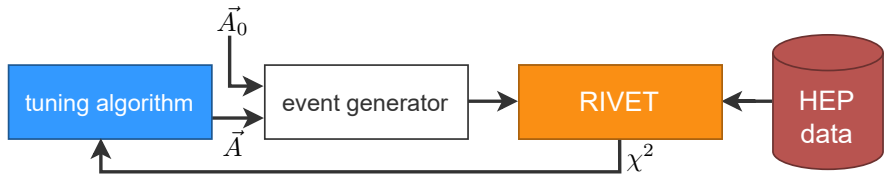
\begin{figure}
    \centering
    \begin{tikzpicture}
        \node[anchor=south west, inner sep=0] (image) at (0,0) {\includegraphics[width=0.8\textwidth]{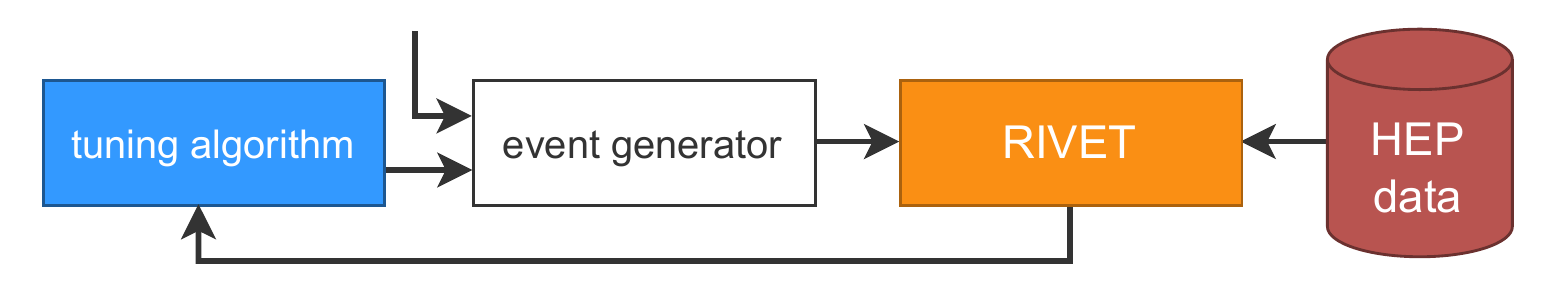}};
        \begin{scope}[x={(image.south east)}, y={(image.north west)}] 
            \node at (0.27,0.98) {\(\vec{A_0}\)};
            \node at (0.27,0.29) {\(\vec{A}\)};
            \node at (0.71,0.205) {\(\chi^2\)};
        \end{scope}
    \end{tikzpicture}
    \caption{A schematic of classic tuning. The classic tuning loop proceeds as follows: 1) A new vector of parameter values $\vec{A_0}$ is used to configure the event generator. 2) The event generator is used to simulate all collision systems fitting the required initial conditions of the \rivet plugins. 3) The \rivet plugins compute predictions comparable to their respective HEP data from the raw event sample. 4) A chi-square value is computed from all \rivet plugins and fed back into the tuning algorithm. 5) The tuning algorithm computes a new vector of parameter values based on this input.} 
    \label{fig:classic tuning}
\end{figure}

The classic tuning loop used in particle physics is illustrated in \autoref{fig:classic tuning}, where a \textit{tuning algorithm} iteratively optimizes the event generator to describe the HEP data. Note that the figure is conceptual only; in practice, tuning adds a few complications to this basic picture. First, one must assign weights to each observable in the calculation of the least-squares-type cost function, because event generators remain incomplete models of nature and usually do not perfectly match measurements to the extent of their implemented physics processes. Some observables are known much more precisely than others, and possibly more precisely than the corresponding theory. This can drag a fit without weights into an implausible parameter space. Introducing weights or imposing additional common model uncertainties is a \textit{ad hoc} way of balancing experimental-versus-theoretical error to avoid this problem. Second, computing the prediction of the event generator is expensive, so it is more economical to construct a \emph{surrogate model} (in the simplest case, a multidimensional polynomial) from the output of the actual event generator, which is used to compute the chi-square instead of the real event generator. Depending on the tuning algorithm, an analytical surrogate model may be needed so compute gradients. 

Examples of automatically tuned models are \sibyll{} and \pythia. While \pythia{} has many tunes, often several per LHC experiment, for example the Monash tune~\cite{Skands:2014pea},  for \sibyll{} automatic tuning was used to determine the best parameters~\cite{Riehn:2019jet}.

Using an interface like \rivet simplifies tuning, but expert knowledge is still needed to select the right measurements that are sensitive to a given range of tuning parameters, to avoid measurements that the event generator was not designed to describe, and to choose between measurements that contradict each other. To facilitate this process, the MCPlots~\cite{Karneyeu:2013aha, Korneeva:2024oho} website has been created. It is based on \rivet and provides comparisons between the established particle physics event generators in many code bases and tunes whose results can be compared to a large set of measurements (\rivet plugins).The website assembles plots of pre-generated results produced on the computers of citizen scientists who contribute to the project via the LHC@home initiative. In addition to the visual comparison of generator predictions with measurements, the website also computes a chi-square test statistic for the goodness-of-fit between the experimental data and an event generator version or tune, allowing non-generator experts to easily select the appropriate tune for their task. 

\subsection{Early tuning efforts for astroparticle physics}
\label{sec:astro-tune}
 
One of the major shortcomings of all current HEP tunes of \pythia~8 is the difficult description of particle collisions at large rapidities, most notably the failure to describe the spectra of neutrons and neutral pions measured at the LHCf experiment~\cite{LHCf:2008lfy}. Specifically, the $\pi^0$ spectra predicted by \pythia~8 are too hard while the neutron spectra are too soft~\cite{LHCf:2017fnw,LHCf:2015nel,LHCf:2015rcj,Fieg:2023kld}. It is worth noting that the event generators used to simulate air showers, such as \eposlhc{}, \qgsjet{}, and \sibyll\ also do not perform well. This is very important for simulating air showers, as well as for physics in the forward direction such as the FASER~\cite{FASER:2023zcr} or SND@LHC~\cite{SNDLHC:2023pun} experiments. 

Starting from different configurations and tunes of \pythia~8~\cite{Christiansen:2015yqa,Skands:2014pea}, a good description of the LHCf data was obtained by adjusting the hadronization model and the parameters of the so-called beam remnants using \rivet and the \textsc{Apprentice} toolkit~\cite{Fieg:2023kld}. Building on the successful ``forward tune" of \pythia~8, efforts are underway to produce a first ``air shower" tune. The aim is to better describe the processes important for air shower simulations, namely the forward hadron production measured e.g.\ by LHCb and NA61~\cite{LHCf:2015nel,LHCf:2015rcj,LHCf:2020hjf,LHCf:2017fnw,Piparo:2023yam,LHCb:2013gmv,LHCb:2016qpe,LHCb:2019avm,LHCb:2021abm,LHCb:2021vww,LHCb:2022dmh,EHSNA22:1990vem,EHSNA22:1991dhh,NA49:2005qor,NA49:2009brx,NA61SHINE:2017fne,NA61SHINE:2017vqs,NA61SHINE:2019aip,NA61SHINE:2022tiz}, as well as the total and inelastic cross sections in different collision systems~\cite{Carroll:1975xf,Carroll:1978vq,Carroll:1978hc,Burq:1982ja,NA22:1987lmr,EHSNA22:1988fqa,}, spanning several orders of magnitude in energy (for the full set of measurements, see~\autoref{tab:rivet_table} in \autoref{app:tuning}).
To achieve this goal, the tuning parameters have to be determined step by step for each collision system, starting with \ppColl, then \pipmpColl/\KpmpColl, followed by \pAColl and \piAColl, and ending with \AAColl. The nucleus $A$ should ideally be of the CNO group, since nitrogen is the most abundant nucleus in the atmosphere, and intermediate-mass nuclei are abundant in the UHECR flux near the spectral cutoff. Therefore, the upcoming proton-oxygen runs at the LHC will provide an important validation data set and future reference point at the TeV scale for this and subsequent tunings.
Parameters related to beam remnants and hadronization are highly relevant for the tuning, as well as the dependence of the parameters on the CM energy of the collision system, since this has to be extrapolated to higher projectile energies. It will also be important to arrive at a tune that describes the mid- and forward-rapidity data without major inconsistencies. 

An essential first step towards global tuning to particle and astroparticle data is to demonstrate the feasibility of tuning to air shower data. To this end, a study based on the event generator \pythia~8 and the air shower simulation code \corsika~8~\cite{Engel:2018akg} was performed by generating mock air shower data using the default settings in \pythia~8. Using Bayesian tuning~\cite{LaCagnina:2023yvi,Schulz:2020ebm} these default values were successfully  recovered by tuning \pythia to \xmax and \nmu in the mock data.
This study shows that tuning to air shower observables is in principle possible, and that the Bayesian tuning approach is a promising method for global tuning to particle and astroparticle data. The experiment also highlights the need for fast air shower simulations. The next step is to perform a first tune to data.

\section{Towards global tuning}
\label{sec:combination}

In this section, we outline a vision for simultaneous, automatic tuning of MC generators with data from particle and astroparticle physics. This requires a framework to be built in collaboration with the developers of the event generators and both experimental communities. 
Many of the necessary ingredients are already available or under development. First studies in this direction have been made with encouraging results~\cite{Fieg:2023kld,Gaudu:2024mkp}.

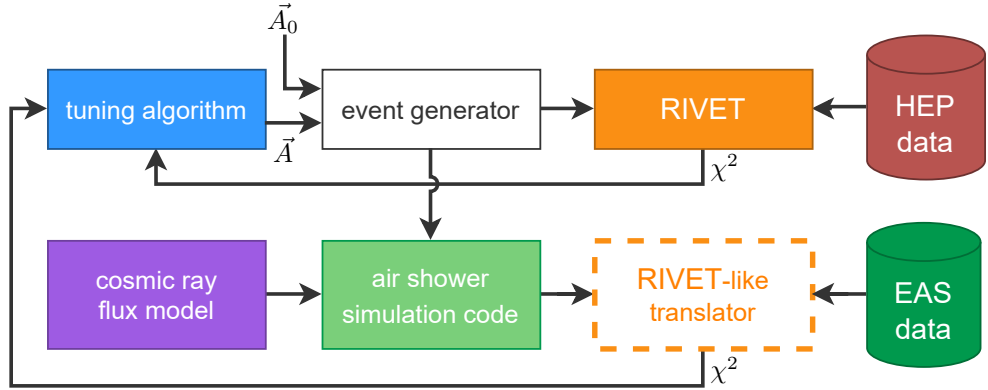
\begin{figure}[th]
    \centering
    \begin{tikzpicture}
        \node[anchor=south west, inner sep=0] (image) at (0,0) {\includegraphics[width=0.9\linewidth]{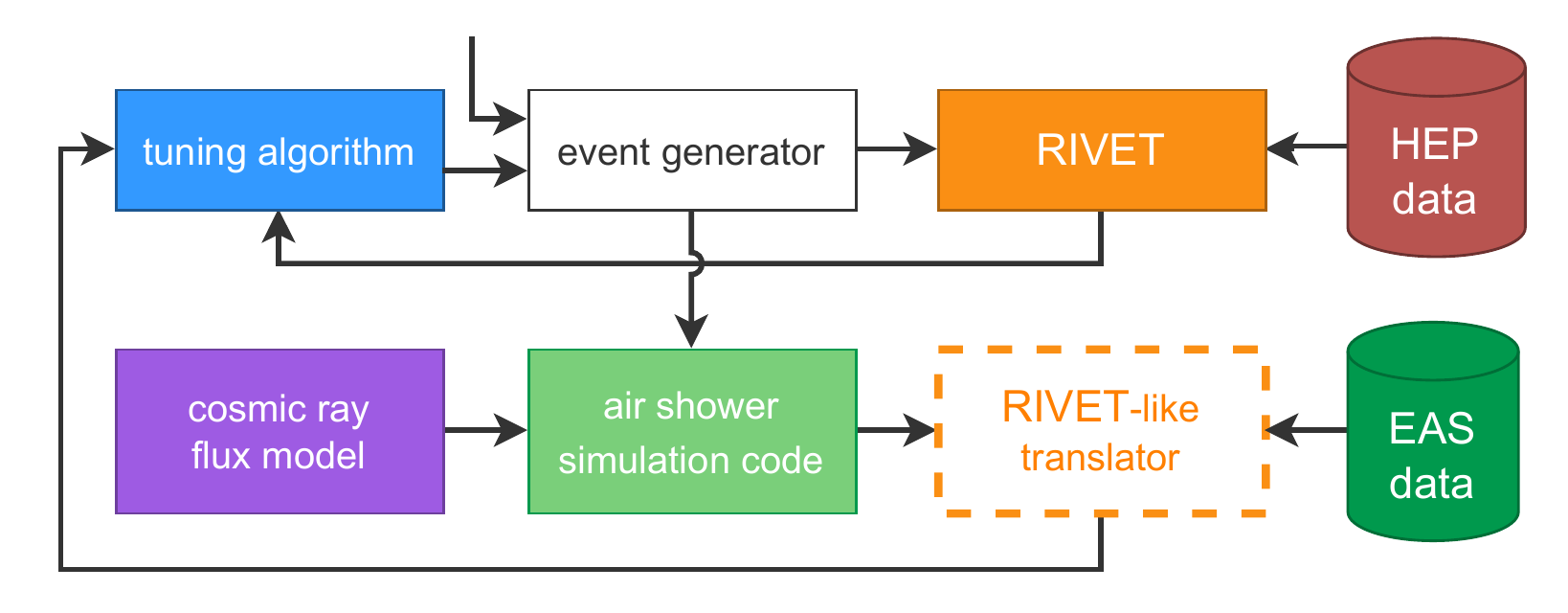}};
        \begin{scope}[x={(image.south east)}, y={(image.north west)}] 
            \node at (0.3,0.98) {\(\vec{A_0}\)};
            \node at (0.3,0.655) {\(\vec{A}\)};
            \node at (0.723,0.61) {\(\chi^2\)};
            \node at (0.723,0.11) {\(\chi^2\)};
        \end{scope}
    \end{tikzpicture}
    \caption{A schematic of global tuning. The global tuning loop proceeds as follows: 1) An initial vector of parameter values $\vec{A_0}$ is used to configure the event generator. 2a) The event generator is used to simulate all collision systems fitting the required initial conditions of the \rivet plugins. 2b) The event generator runs inside the air shower simulation code \corsika{} to simulate air showers, or simulates all collision systems required to build interaction tables for \conex{} or \mceq{}, which are then used to simulate air showers. 3) The energy spectrum and mass composition of the primary cosmic rays follow a cosmic ray flux model. 4a) The \rivet plugins compute predictions comparable to their respective HEP data from the raw HEP event sample. 4b) \rivet-like translators compute predictions comparable to their respective EAS data based on the raw air shower sample. 5) A chi-square value is computed from all \rivet plugins and the translator plugins and fed back into the tuning algorithm. 6) The tuning algorithm computes a new vector of parameter values based on these inputs. Note that the \rivet-like translator still needs to be developed.}
    \label{fig:global tuning}
\end{figure}

The basic training loop as envisioned for a global tuning is illustrated in \autoref{fig:global tuning}. The particle physics tuning loop as discussed in \autoref{fig:classic tuning} is complemented by a second iterative loop where the input from the particle physics event generator and the cosmic ray flux model are combined in the  air shower simulation code to derive the prediction for the observables that is to be compared with the air shower data. The task of the Rivet-like translator is to choose the appropriate configuration for the air shower simulation for each observable. This includes selecting the method of calculation, which can be a full MC simulation (\corsika{}) or solving cascade equations (\conex{}, \mceq{}). 

The data sets from astroparticle and accelerator based measurements are largely complementary, many EAS observables are known to better than 10\%, which together with the lever arm in phase space provides significant constraints on event generators. This is also demonstrated by the fact that some EAS observables are difficult to describe by current event generators.

\subsection{Current challenges in describing EAS data}
\label{sec:tuning challenges}

The consistent description of EAS data is a challenge for current event generators.  They show a muon deficit compared to measurements~\cite{PierreAuger:2014ucz,PierreAuger:2021qsd}. The muon production depth \xmumax is also not reproduced by all event generators: \eposlhc{} predicts an excessive depth in simulations, while \qgsjet{}-II.04 is consistent~\cite{PierreAuger:2014zay}. 

\paragraph{Strangeness enhancement}
The influence of enhanced strangeness production on muon production in air showers has been studied in high multiplicity events with a core-corona model, where the core is responsible for the strangeness enhancement~\cite{Baur:2019cpv}. The studies were based on a modified version of \eposlhc{}. Using the $\Omega/\pi$ ratio measured by ALICE as a function of hadron multiplicity as a reference and assuming that strangeness production for all collision systems depends only on the hadron multiplicity, the LHC results were extrapolated with the core-corona model to ultra-high energy air showers. It was found that this leads to an increase in the muon number of about 10\% , which is not sufficient  to resolve the observed muon deficit in air shower simulations.

\paragraph{Standard Model uncertainties}
The impact of Standard Model uncertainties on the predictions of the muon number \nmu and the muon production depth \xmumax has been studied in~\cite{Ostapchenko:2024xbe}. The EAS observables were computed with \conex{} and the event generators \qgsjet{}-II-04, \eposlhc{}, \sibyll-2.3, and a modified variant of \qgsjet{}-III. Modifications of the latter were made within experimental constraints from collider data. The influence of the following modifications was studied: a change in the pion energy distribution, an enhanced gluon content of the pion, a modified energy dependence of pion exchange, and a possible enhancement of (anti)nucleon, kaon and meson production. It was found that all these variations cannot increase \nmu by more than 10\%, which is not enough to agree with the data from the Pierre Auger Observatory. Moreover, at the same time they further increase \xmumax which aggravates the tension with the Pierre Auger Observatory observations.

\paragraph{\sibyll$^{\bigstar}$}
A similar study was performed with a modified version of \sibyll-2.3d, called \sibyll$^{\bigstar}$~\cite{Riehn:2023wdi}, with \textit{ad hoc} modifications of the final state ignoring internal model consistency.
Simulations were performed to study different scenarios for increasing muon production in air showers: increased baryon (and hyperon) production, increased $\rho^0$ production, increased K$^\pm$ production, and a mixture of baryon and $\rho^0$ production. It was found that all scenarios can match the Auger data, but the scenarios with increased $\rho^0$ do so without introducing sudden jumps in the production cross section as a function of the collision energy. However, the forward/high $x_F$ $\rho^0$ production cross section is expected to decrease with increasing energy, while it remains constant or even increases in the modified scenarios that match Auger data.

\paragraph{Strangeball model}
The muon production in air showers can be increased with a so-called strangeball model~\cite{Manshanden:2022hgf}, an evolution of the fireball model of~\cite{Anchordoqui:2016oxy}. The latter assumes that a quark-gluon plasma droplet (the fireball) is formed in a fraction of the collisions. The fireball and core-corona models are similar, but in the latter both the core (fireball) and the corona (string fragmentation) contribute simultaneously, whereas a fireball is formed only with some probability but then is the sole product of the collision. The authors found that the fireball model cannot consistently describe the mean and variance of \xmax in air showers due to the inelasticity enhancement associated with the formation of a plasma state. This problem does not arise in core-corona models, where the core component does not extend to very forward rapidities, so that its effect on inelasticity is reduced. The strangeball model solves this problem by increasing only the strangeness produced in Standard Model hadronic interactions relative to other flavors, without forming a quark gluon plasma state. Strangeball parameters have been found that are consistent with the muon production seen in Auger data. Constraints from measured shower-to-shower fluctuations of the muon number require strangeness enhancements already at the TeV scale. A comparison with relevant measurements from the LHCf and LHCb detectors does not directly exclude this scenario. Further LHCb measurements in Run 3 and Run 4 at the LHC could further constrain this model.

\vspace{2mm}
Global tuning may reveal that current event generators are not able to consistently describe all the data, and that our current modeling of non-perturbative QCD is incomplete. This would be a success of global tuning, as it would help to guide us towards the appropriate extensions of the standard theory. 

\subsection{A RIVET-like translator for astroparticle data} 
\label{sec:rivet extension}

Global tuning is necessarily an automatic tuning that adjusts many parameters in the event generator to many data sets simultaneously. When tuning only to HEP data, it is possible to select data sets that correspond to only a small subset of parameters in the event generator and perform a manual tuning. This approach is not feasible when including EAS data which are sensitive to many correlated parameters. For EAS measurements, a translator software analogous to \rivet discussed in \autoref{sec:autoTuning}, or an extension of the \rivet software is needed to allow global tuning. It seems obvious that this effort should build on the existing infrastructure developed within the HEP community. 

A translator for EAS data can work in the same way as a \rivet plugin, but with simulated air showers as input. For each analysis, one defines the relevant energy range of air showers to be simulated, the atmospheric profile, the zenith angle range, and the observation level. Air showers are then simulated according to these specifications. The translator applies any selections that bias the air-shower sample, such as cuts on the depth of shower maximum \xmax that correspond to the field of view of the fluorescence telescopes.

The existing \rivet software can be extended to become this new translator, but this is not a trivial task. Several arguments speak in favor of this approach: The internal \rivet file format is flexible enough to describe EAS data as well. Tuning software can already interface with \rivet plugins, integrating EAS plugins would be relatively easy, and the MCPlots website could be used to show comparisons with EAS data. The main problem is that \rivet is designed for input in the form of a particle graph, which cannot represent air showers. An air shower consists of longitudinal density profiles for electrons, photons, and muons, and a list of millions of particles that reach the observation level. The HepMC format is not designed to represent density profiles, nor to store millions of particles. Handling this input would likely require deep changes to \rivet, which would need to be discussed with the \rivet developers to evaluate the feasibility of such changes. 

The alternative is to construct a new \rivet-like translator from scratch. For this approach, tuning and plotting software would have to be adapted to interface with this new translator. 
An advantage would be the decoupling from the \rivet release cycle, and the possibility to write the translator in Python instead of C++ to benefit from the fast development cycle and easy prototyping. 

\paragraph{Tuning without a surrogate model}
One limitation of established automatic tuning methods is the need to construct a surrogate model. This step adds complexity because the parameter space must be efficiently sampled. Methods based on stochastic gradient descent (SGD), such as \textsc{Adam} \cite{Kingma:2014vow}, could potentially be used to tune the event generator directly, avoiding the construction of the surrogate model, and dramatically reducing the computational cost of tuning many parameters at once.

\subsection{Tuning of the cosmic ray flux model} 
\label{sec:tuning cosmic ray flux}

An optional but natural extension of global tuning is the inclusion of cosmic ray flux model parameters, which provides additional benefits. Air shower observables depend on the cosmic ray flux and its composition, which are \apriori unknown and must be inferred from these observables. The latest generation of hybrid observatories measure the all-particle cosmic ray flux in a nearly model-independent way, which is now fairly well understood~\cite{Dembinski:2017zsh}. The remaining uncertainty in the all-particle flux is dominated by the energy scale uncertainty, which has little effect on most air shower observables of interest. Exceptions are the atmospheric muon and neutrino fluxes. The composition, however, carries large uncertainties, and affects several air shower observables.

This seems to introduce a circular dependency: we want to use air shower observables to tune event generators, but the prediction of these observables depends on the composition that we infer by comparing event generator predictions with air shower observables. In reality, this is not a circular dependency, because we have many observables that are sensitive to different aspects of the event generators. The composition is usually inferred from the depth of the shower maximum, \xmax, and predictions based on this are then compared with other observables. This is successful, because the theoretical uncertainty in predicting \xmax is smaller than for other variables.

An alternative is to include a cosmic ray composition model in the tuning process. The parameters of the flux model are then also adjusted by the tuning algorithm.
This \emph{universal tuning} would infer the composition from all air shower observables, not just from \xmax. This approach can only work if the tuned event generator is able to consistently describe all air shower observables. This is not currently the case, and naive universal tuning would lead to incorrect results. However, the approach can still be used if \xmax measurements are given a higher weight in the least-squares-type cost function used in the tuning.

In addition to conceptual consistency, the benefit of universal tuning is an additional result: the model of the cosmic ray composition obtained in this way would have an uncertainty band propagated from all air shower and HEP observables used in the tuning. Currently, the uncertainty is estimated from the scatter of results obtained with different event generators. This uncertainty estimate may over- or underestimate the true uncertainty.

\section{Summary and next steps}
\label{sec:summary}

In this article, we demonstrated that a global tuning of event generators with data from both high-energy accelerators (HEP) and extensive air showers (EAS) has the potential to resolve some of the puzzles we currently face in interpreting astrophysical measurements. 
As an example, we discussed effects that could increase the muon production in air showers to better describe observations. 
Global tuning will either resolve such tensions or lead us to an extension of the standard theory. 

Global tuning needs to be done with automatic tuning software for which the classic tuning loop is extended by including a model of the cosmic ray flux and air shower simulations. To facilitate the use of EAS data, a translator similar to \rivet in HEP data needs to be developed. This report can serve as a starting point for such a development. 

Another requirement for global tuning is an event generator with a documented tuning interface that includes all the physical processes needed to simulate hadron-nucleus and nucleus-nucleus collisions up to hundreds of TeV. \pythia~8/Angantyr currently best meets these requirements and is therefore the first target for tuning, but other event generators should follow.

Established methods for tuning are ready to be used for global tuning, as demonstrated in toy studies. One concern is the significant additional computational cost of running air shower simulations. This cost can be significantly reduced by using cascade equation solvers such as \conex{} and \mceq{} instead of full \corsika{} simulations. For tuning of many parameters at once, future methods based on stochastic gradient descent may perform better, but such methods have yet to be developed. In the near future, air shower simulations will also benefit from ongoing efforts to improve the description of forward hadron production in event generators and from data using oxygen beams in the LHC.

In summary, global tuning is within reach. Prototypes and smaller studies have shown that the tuning process works and is useful to address the main challenges in simulating EAS and HEP data. With the combined efforts of the community, we can make global tuning into a standard tool that everyone can use. Global tuning can lead to improvements in tuning methods themselves, benefiting the (astro)particle physics community as a whole. We can expect many interesting results in the near and midterm future.

\clearpage\newpage 
\section*{Acknowledgments}
This paper is a comprehensive overview of work that has been advanced with a collaboration of experts during the workshop \emph{Tuning of hadronic interaction models}\footnote{\url{https://indico.uni-wuppertal.de/event/284/}} at the Bergische Universität Wuppertal in January 2024. The international workshop was organized as part of the Collaborative Research Center SFB1491, \emph{Cosmic Interacting Matters --- From Source to Signal}. We acknowledge the support of the workshop and the related research by many of the workshop participants and authors of this paper by SFB1491, funded by the Deutsche Forschungsgemeinschaft (DFG, German Research Foundation) under project number 445052434.

J.~Albrecht acknowledges additional support from the Heisenberg Programme of the Deutsche Forschungsgemeinschaft (DFG, German Research Foundation) under project number AL 1639/5-1, and from the Bundesministerium für Bildung und Forschung (BMBF, Federal Ministry of Education and Research) under grant number 05H21PECL1 within ErUM-FSP T04.
H.~Dembinski acknowledges funding from the DFG under project number 449728698. 
Karl-Heinz Kampert acknowledges additional support from the BMBF under grant numbers 05A20PX1 and 05A23PX1 and from DFG under project no.\ 445990517. G.~Sigl acknowledges support from the DFG under Germany’s Excellence Strategy --- EXC 2121 \emph{Quantum Universe} --- 390833306, and from the BMBF under grants 05A20GU2 and 05A23GU3.
N.~Korneeva acknowledges support from the Monash Warwick Alliance as part of the Monash Warwick Alliance in Particle Physics, and from the LHC Physics Centre at CERN (LPCC).
S.~Ostapchenko acknowledges support from the DFG under project numbers 465275045 and 550225003.
F.~Riehn has received funding from the European Union’s Horizon 2020 research and innovation programme under the Marie Skłodowska-Curie grant agreement number 101065027.
T.~Sjöstrand has been supported by the Swedish Research Council under contract number 2016-05996.
L.~Cazon thanks Ministerio de Ciencia e Innovacion / Agencia Estatal de Investigacon (PID2022-140510NB-I00 and RYC2019-027017-I), Xunta de Galicia (CIGUS Network of Research Centers, Consolidacion 2021GRCGI-2033, ED431C-2021/22 and ED431F-2022/15), and the European Union (ERDF).
J.~Blazek, J.~Ebr and J.~Vícha have received funding from the following grants: CAS LQ100102401, GACR 21-02226M and MEYS CZ.02.01.01/00/22\_008/0004632.
P. Paakkinen acknowledges the support from the the Research Council of Finland (projects 330448 and 331545) and as a part of the Center of Excellence in Quark Matter of the Research Council of Finland (project 364194).

\appendix

\section{Details on event generators }
\label{app:had-models}
\subsection{EPOS4}

The theoretical basis of \epos{}4\footnote{Available at \url{https://klaus.pages.in2p3.fr/epos4/}}~\cite{Werner:2023zvo,Werner:2023fne,Werner:2023mod,Werner:2023jps} is the so-called parton-based Gribov-Regge field theory (GRFT) \cite{Werner:2005jf} with energy conservation at amplitude level and a bare amplitude based on the parton model. The main new development in \epos{}4 is a way to accommodate simultaneously: rigorous parallel scattering, energy-momentum sharing, perturbative QCD, and validity of the Abramovsky-Gribov-Kancheli (AGK) theorem \cite{Abramovsky:1973fm}. This ensures binary scaling (in \AAColl scattering) and factorisation (in \ppColl) for hard processes, by introducing saturation (in a particular way), compatible with recent \enquote{low-$x$-physics} considerations. Energy-momentum sharing is mandatory for a consistent picture and offers a connection between factorisation and saturation.

The validity of AGK means that one can do the same as models based on the QCD factorisation theorem to study inclusive cross sections, and much more. As mentioned in the introduction, many important observables go beyond inclusive cross sections, where one needs access to complete events. The unique feature of \epos{}4 is to start from a parallel multiple scattering scenario, while ensuring that it breaks down to the description using PDFs for inclusive cross sections. Other models start with inclusive cross sections and then introduce multiple scattering via the eikonal formula. This is relevant also for nuclear collisions, where the same strategy is used of starting with parallel scatterings without ordering (a unique feature). Doing nuclear scattering in \epos{}4 is a natural extension of the \ppColl approach.

From the initial stage of parallel instantaneous partonic scatterings, a number of pre-hadrons are obtained. In the core-corona approach, the pre-hadrons are separated into \enquote{core} and \enquote{corona} pre-hadrons, depending on the energy loss of each pre-hadron when traversing the \enquote{matter} composed of all the others. Corona pre-hadrons (per definition) can escape and become final hadrons, whereas core pre-hadrons lose all their energy and constitute what we call \enquote{core}, which acts as an initial condition for a hydrodynamical evolution of a quark gluon plasma.

The evolution of the core ends when the energy density falls below a critical value and hadrons are formed. \epos{}4 uses a new procedure of energy-momentum flow through the \enquote{freeze-out hypersurface} defined by the critical energy density value, which allows for defining an effective invariant mass. The latter decays according to microcanonical phase space into hadrons, which are then Lorentz boosted according to the flow velocities computed at the hypersurface. New efficient methods for the microcanonical procedure were developed to make this feasible. Energy-momentum and flavours are conserved in the full scheme for all hadrons from core and corona.

\subsection{EPOS LHC-R}

\eposlhc{}-R\footnote{Not yet publicly available.}~\cite{Pierog:2023ahq} is an updated and re-tuned version of \eposlhc{}\cite{Pierog:2013ria}. As \epos{}4, it is based on the parton-based GRFT with energy conservation at amplitude level, but without the new saturation treatment. A custom model for the parton distributions is used, the valence quarks are based on the Glück-Reya-Vogt (GRV) parameterisation \cite{Gluck:1998xa}. It also uses the core-corona approach. The calculation of the hadronisation in high energy/density environments, such as high-multiplicity events or heavy ion collisions, is implemented in a simplified manner compared to \epos{}4. Instead of running a full hydrodynamical evolution of the core, collective flow is calculated from a parameterisation, which is matched to the full simulation. These simplifications speed up the simulation of events, which is an important feature for the computationally expensive EAS simulation, which is dominated by the computation time consumed by the hadronic event generator.

\eposlhc{}-R is tuned to newer and more detailed data sets than \eposlhc, which was released in 2012, when only early LHC data was available. This lead to a lower extrapolated cross section and better pseudorapidity distributions. Some missing physics phenomena have been introduced, such as colour transparency, for a better description of multiplicity fluctuations in nuclear interactions for instance, or a real pion exchange process for very forward neutron production as measured by the LHCf experiment. Special attention is given to the details of the hadronisation chain, in particular on the $\rho$ meson production in string fragmentation and how the energy is shared between the core and corona contributions. \eposlhc{}-R is meant to be a transition between \eposlhc and \epos{}4, which is much further evolved and includes new features like saturation effects.

\subsection{QGSJET-III}

QGSJET-III\footnote{Code is available on request.}~\cite{Ostapchenko:2024jsg,Ostapchenko:2024myl} is a recent update of QGSJET-II-04~\cite{Ostapchenko:2010vb,Ostapchenko:2013pia}, from which it inherits its principal feature: a microscopic treatment of nonlinear interaction effects, based on all-order resummation of the underlying Pomeron-Pomeron interaction diagrams. The evolution of parton densities is calculated according to the DGLAP equations \cite{Altarelli:1977zs,Dokshitzer:1977sg,Gribov:1972ri}. The new theoretical mechanism implemented in QGSJET-III is the treatment of higher twist (HT) corrections to hard parton scattering processes, based on a phenomenological extrapolation of the corresponding approach of Qiu and Vitev \cite{Qiu:2003vd,Qiu:2004da}, regarding deep inelastic and proton scattering on heavy nuclei, towards collisions of hadrons with protons and light nuclei. While the magnitude of such HT corrections is strongly constrained by HERA data on deep inelastic electron--proton scattering, such that their impact on the predicted cross sections and particle production yields in \ppColl and hadron-air collisions is rather moderate, the mechanism does its principal job: taming the rise of inclusive cross sections for (mini)jet production, in the limit of small jet transverse momenta \cite{Ostapchenko:2024jsg}.

Besides that, an important technical improvement implemented in QGSJET-III concerns the treatment of the pion exchange process in hadron--proton and hadron--nucleus (nucleus--nucleus) collisions \cite{Ostapchenko:2024myl}. This mechanism strongly impacts predictions for the muon content of extensive air showers (EAS) initiated by primary cosmic rays \cite{Ostapchenko:2013pia}. Here the process is described in a theoretically more consistent way, compared with the simplified treatment of QGSJET-II-04, restricted to pion--proton and pion--nucleus collisions only. Further, data on forward neutron production from the NA49 \cite{NA49:2009brx} and LHCf \cite{LHCf:2018gbv} experiments have been used to verify the validity of the approach, regarding the energy dependence of the process, over six decades in energy.

Applying the QGSJET-III model to EAS simulations, one obtains rather small differences with respect to QGSJET-II-04 for basic EAS characteristics, like the shower maximum depth \xmax or the muon number \nmu at ground level \cite{Ostapchenko:2024myl}. This may indicate that the treatment of relevant aspects of hadronic interaction physics is already sufficiently constrained by available accelerator data. Such a conclusion is further corroborated by a recent study, where it has been demonstrated that the predicted EAS muon content can not be increased by more than 10\% while staying within the standard physics picture (which here excludes core-corona and similar approaches), without entering into a serious contradiction with relevant accelerator data \cite{Ostapchenko:2024xbe}.

\subsection{Sibyll}

\sibyll{}\footnote{Available in Chromo~\cite{Dembinski:2023esa} and CRMC~\cite{2021zndo...5270381U}, \corsika{} \cite{Heck:1998vt,Engel:2018akg,CORSIKA:2023jyz}, and all air shower codes. Code and source documentation are available upon request.}\cite{Ahn:2009wx,Riehn:2019jet,Fedynitch:2018cbl} is designed to describe the general features of hadronic multi-particle production, like the leading-particle effect, the formation of high-$\pT$ jets predicted in QCD, the production of diffractively excited states of the projectile and target, and approximate scaling of leading-particle distributions with interaction energy. Focus is put on those physics aspects that are most relevant for the development of extensive air showers, like energy flow and particle production in the forward phase space region. While the model is kept as simple as possible, the important microscopic physics concepts and the general principles of scattering theory and unitarity are implemented to allow for extrapolation to energies and phase space regions beyond the reach of colliders.

The interaction model in \sibyll{} is based on the two-component dual parton model with soft and hard minijets~\cite{Capella:1992yb}. Nuclear collisions are treated with an extended superposition model~\cite{Engel:1992vf}. It also includes low- and high-mass diffraction and a model for the excitation of beam remnants~\cite{Drescher:2007hc}. Hard scattering is distinguished from soft scattering by an energy-dependent cut-off $\pTo$ in transverse momentum. The cross section for hard scattering is calculated to leading order (LO) in QCD at the scale $\pTo$, for soft scattering a parameterisation based on the Regge field theory~\cite{Donnachie:1992ny} is used. The energy evolution of parton densities and saturation is effectively included by the increase of $\pTo$ with energy. Contributions from quarks of all flavours and gluons are included in the QCD cross section, in the subsequent fragmentation (based on the Lund model) only hadrons containing $(u,d,s)$ and $c$ quarks are produced.

\subsection{P{\small YTHIA}~8}

\pythia~8\footnote{Available at \url{https://pythia.org/}, a detailed (HTML) manual is distributed with the code.}~\cite{Bierlich:2022pfr} is a general-purpose event generator. At the LHC it is primarily used to simulate soft and hard processes in the central rapidity range of \ppColl collisions. An MPI (multiparton interaction) is the basic building block of an event. It involves a perturbative $2 \to 2$ QCD process, where the $\pTo$ parameter is introduced to dampen the low-$\pT$ divergence. A varying impact parameter, with a non-zero Poissonian number of MPIs at each, builds up the MPI multiplicity spectrum. This is similar to but not equivalent with traditional Gribov--Regge. The $\pT$-ordered initial- and final-state showers are added to each MPI.

The Lund string fragmentation model is used to hadronise events. Naively, each MPI would have strings stretching out to the beam remnants, but colour reconnection is applied to reduce the total string length, both to the beam remnant and in the central region itself. Diffractive events are handled as if a Pomeron acts like a glueball \cite{Ingelman:1984ns}.

In recent years, \pythia~8 has been extended \cite{Sjostrand:2021dal} to enable its use in EAS simulation. PDFs were introduced for more than 20 different hadron species. Total and partial cross sections were introduced for corresponding hadron--nucleon collisions. These were smoothly matched onto the low-energy hadronic rescattering framework~\cite{Sjostrand:2020gyg}. Thus, \pythia can be used almost from threshold energies up to and beyond centre-of-mass energies of 100 TeV, aside from increasing problems with numerical precision. A simple kludge was introduced to handle nuclear targets, tuned to approximately reproduce the \textsc{Angantyr} (see below) results at collider energies. As a non-standard feature, not fully available in the public version, rope hadronisation~\cite{Bierlich:2014xba} and shoving~\cite{Bierlich:2016vgw} can be used to enhance the fraction of strange baryons and to generate asymmetric flow, respectively.

\textsc{Angantyr}~\cite{Bierlich:2018xfw} is a module of \pythia~8, that extends its functionality from \ppColl to \pAColl and \AAColl. It gives a special attention to fluctuations in the nucleon wave function. The Good--Walker formalism \cite{Good:1960ba} is used to obtain the relevant cross sections. Glauber modelling is used to find which nucleons collide. Then, all non-diffractive sub-collisions are considered in order of increasing impact parameter. Altogether, a reasonable description is obtained for multiplicity and pseudorapidity distributions. \textsc{Angantyr} should be able to offer a better alternative for cosmic ray evolution than the current kludge, and is being tried out.

\subsection{UrQMD}

The UrQMD\footnote{Available at \url{www.urqmd.org}.}~\cite{Bass:1998ca,Bleicher:1999xi} (Ultra-relativistic Quantum Molecular Dynamics) model is based on the propagation of hadrons according to Hamilton's equation of motion as derived from the Ritz variational principle. The model allows to obtain the full time evolution of the collision, from the initial state to the final particles. The decoupling of the particles from the system is governed by their individual scatterings. The propagation uses relativistic kinematics and includes, as usual for QMD-type simulations, all $n$-particle correlations. This is different from test-particle based approaches that solve certain kinds of Boltzmann-type equations and therefore only allow 1-particle distribution functions to be obtained (meaning that they do not propagate particle correlations). UrQMD includes potential interactions of Skyrme type (but other potentials can also be included), and allows using soft and hard potential interactions. The current data suggest that hard potentials, meaning a stiff equation-of-state, is appropriate at low energies ($\sqrt{s_{NN}} < 7$ GeV), while a soft equation-of-state is favoured at higher collision energies \cite{Hillmann:2018nmd}. The hadrons do interact according to a 2-particle collision term that may involve elastic interactions and inelastic reactions, where the inelastic reactions include resonance creation and decay, and also string formation and fragmentation. At very high energies ($\sqrt{s_{NN}} > 50$ GeV), the model employs \pythia to include hard scatterings. The table of included hadrons reaches typically up to 3~GeV in mass, depending on particle type. In the latest version also charm degrees of freedom have been included, although charm production has only been benchmarked in the low energy range ($\sqrt{s_{NN}} < 7$ GeV). Generally, the application of such a model set-up leads to a rather natural transition from the central collision area to the periphery of the interaction zone. That is often termed as the core--corona transition, as it leads to substantial thermalization in the central region of the collision due to high interaction rates, and then naturally transforms into single nucleon+nucleon interactions towards the edges of the collision zone. For a recent review of transport models the reader is referred to Ref.~\cite{Bleicher:2022kcu}.
\section{Transport codes }
\label{app:transport}

\subsection{CORSIKA}

\corsika{} is a Monte Carlo code for the simulation of extensive air showers, that was originally developed for the KASCADE experiment but has found wide application in many astroparticle physics experiments~\cite{Heck:1998vt}. \corsika{} is a monolithic Fortran code, which exhibits excellent performance but incurs limitations such as limited parallelization possibilities and increasingly difficult maintenance. To address these problems and leverage the possibilities of modern software engineering, since 2018 a rewrite of \corsika{} in modern C++17 is carried out, with a focus on modularity and the possibilities and needs of modern supercomputing~\cite{Engel:2018akg,Alameddine:2024cyd}. This new version, called \corsika{}~8, is now physics-complete and offers unprecedented flexibility in air shower simulations~\cite{CORSIKA:2023jyz}, including radio and Cherenkov emission. In comparison to the highly-optimized but more specialised preceding Fortran versions, \corsika{}~8 is still slower by a factor 3--5, but offers new possibilities such as complete genealogy of air shower particles; showers in different media, \eg crossing from the atmosphere into ice or rock; multithreaded radio emission calculation; and GPU-parallelised calculation of the Cherenkov emission of air showers.

One of the main applications of \corsika{} concerns the simulation of EAS and their comparison with observations.
In particular, measurements of the muon content of UHECR-induced EAS, notably by  the Pierre Auger Observatory, show that hadronic multi-particle production in EAS is not yet fully understood \cite{PierreAuger:2016nfk,PierreAuger:2024neu}. Inspecting genealogical information (such as generation
number, particle species of its preceding generations, etc.) of the particles in EAS simulations can yield valuable insight \eg into the production mechanism of muons and the gradual decoupling and evolution of the electromagnetic cascade. Moreover, it is possible to make quantitative statements that can be compared with Heitler-Matthews-like toy models \cite{Matthews:2005sd} that describe EAS observables qualitatively. In particular, the number of hadron generations that precede the decay into a muon has been studied as a function of primary energy, zenith angle and lateral distance, and the muon production depth profiles as a function of hadron generation \cite{Reininghaus:2021zge}. Furthermore, the EM profiles as a function of hadron generation have been investigated, and it has been worked out to what extent $X_{\rm max}$ is determined already by the primary interaction alone and which influence later hadron generations exert \cite{Ulrich:2010rg}.

\subsection{CONEX}

\conex{} \cite{Bergmann:2006yz} is a Fortran code which was designed to realistically simulate the longitudinal air shower development for experiments that observe air showers with fluorescence or air-Cherenkov telescopes. \conex{} can simulate the longitudinal profile much faster than a full Monte-Carlo approach, and can do so without employing the thinning technique that is mandatory for the simulation of ultra-high energy air showers. It is also commonly used in studies that explore the impact of modified hadronic interactions on air shower observables.

\conex{} uses a hybrid approach. The first steps of the hadronic cascade, which dominate its shower-to-shower fluctuations, are computed with the Monte-Carlo technique. Afterwards, particles are binned, and the cascade equations solved numerically. That way, \conex{} can also be used to simulate observables related to shower-to-shower fluctuations of the depth of shower maximum \xmax or the muon number \nmu. \conex{} was later integrated in \corsika{}~7 \cite{Pierog:2011kzx}, which allows one to perform 3D hybrid calculations, where the last stage of the air shower is again simulated with the Monte-Carlo technique. This speeds up air shower simulation by a factor of five compared to the full Monte-Carlo approach with thinning at high energies. In most studies, however, \conex{} is used in its original mode.

\subsection{MCEq}

\mceq{} \cite{Kozynets:2023tsv} is an open-source numerical cascade-equation solver written in Python that has been developed and optimized for the calculation of atmospheric lepton fluxes. It is conceptually similar to \conex{}, but does not use Monte-Carlo simulation for the initial stages of the shower. It allows one to use various hadronic interaction models, among them \sibyll, \eposlhc{}, QGSJet and \dpmjet{}. For underground transport the \proposal{} code has been established. Results obtained with \conex{} and \mceq{} are numerically similar, if the same event generators are used. Models generally show a muon flux deficit of 30-35\% above a few tens of GeV. Although \mceq{} is written in Python, its computational speed exceeds that of \conex{} (a few seconds versus tens of minutes), since the numerical heavy-lifting is done by fast third-party libraries that allow one to exploit the sparsity in the system of equations.

Atmospheric neutrino fluxes are of particular interest for neutrino experiments such as IceCube as they represent a foreground to the astrophysical neutrino fluxes. Atmospheric neutrino fluxes are divided into a conventional component resulting from the decay of pions and kaons while the so-called prompt neutrinos result from the decay of heavy mesons and other hadrons; the largest contribution to the prompt flux is coming from the decay of D mesons. Due to relativistic time dilatation of long-lived light mesons at high energies, the spectrum of the conventional neutrinos is steeper than the primary cosmic ray flux by about $1/E$, whereas the prompt ones roughly follow the primary fluxes. Because of this difference in spectral slopes, the prompt component starts to dominate the atmospheric lepton flux above 10 PeV.

\subsection{CRPropa}

In contrast to the other transport codes, \crpropa{} is designed to model the transport of high-energy particles over galactic, intergalactic, and cosmological distances. The focus is on movement in coherent and turbulent magnetic fields, and photo-nuclear interactions with cosmic photon background fields. The recently released version \crpropa{} 3.2 \cite{AlvesBatista:2022vem} is the latest update in a continued effort to maintain and extend this open-source code well known in the cosmic-ray community. Originally aimed at simulating the ballistic propagation and interactions of Ultra-High Energy Cosmic Rays \cite{Armengaud:2006fx,Kampert:2012fi}, today it can handle diffusive propagation of cosmic rays in a variety of magnetic fields \cite{Merten:2017mgk}, deal with stochastic cosmic ray acceleration, model electromagnetic cascades for gamma ray emission and transport, among other capabilities. It was possible to include diffusive propagation into the existing framework by representing particle densities as pseudo-particles whose trajectories can be simulated like ordinary particles.

It is currently not expected that \crpropa{} can be used for event generator tuning, but this may change in the future. The latest public \crpropa{} version does not simulate hadronic interactions of cosmic rays with interstellar gas, so event generators have no impact on current \crpropa{} predictions. However, work is currently ongoing to implement these interactions either by directly calling event generators or by using precomputed tables \cite{Morejon:2023zbw}. It might then become possible to use neutrino and gamma emissions from strong sources for tuning, as neutrinos are produced by decays of charged pions, while gamma rays are produced by decays of neutral pions.

It is worthwhile to highlight the mathematical and technological approaches used in \crpropa{}, which partially inspired other transport codes, in particular, \corsika{} 8. \crpropa{} is written in a mix of C++ and Python, uses a modern modular design, and features sophisticated parallelization techniques. In \crpropa{}, each step in the simulation is handled by a module, which implements a physical process. Each module processes a stack of (pseudo)particles at once, which allows one to use vectorization capabilities of modern CPUs and GPUs. \crpropa{}'s design offers a great deal of flexibility for debugging, prototyping, and experimentation. It is possible to write modules in pure Python or in C++. In most frameworks, the main loop that passes the particle stack from module to module is written in C++ and cannot be directly accessed from the Python layer. In \crpropa{}, one can replace the standard loop completely with a plain Python loop, which provides a maximum of control for experts and introspection. The high latency of executing Python calls is not an issue in this approach, if enough particles are placed on the stack. Then, the overall computation time is still dominated by the time spent in each module, which makes this approach feasible.

\subsection{Z-moment method}

Atmospheric lepton fluxes can be computed to good approximation with the semi-analytical Z-moment method (see \eg \cite{Gaisser:2016uoy}). In this approach, the cascade equations are solved for a continuous power-law energy spectrum of cosmic rays under several assumptions, for example, superposition for the projectile (as far as forward production is concerned, a projectile nucleus can be treated like a superposition of its nucleons, each with the same fraction of the total energy). A Z-moment is the result of the integral over the input nucleon spectrum and the (energy-)differential cross section for the production of the desired lepton. While this method introduces approximations which have to be checked against numerical codes, it is very transparent and thus useful to study the impact of uncertainties in cross sections on the atmospheric lepton flux. It cannot be used to simulate monoenergetic air showers, however, since the assumption of a continuous input spectrum is baked into the method.

The theoretical uncertainty of the prompt atmospheric neutrino flux was studied with Z-moments in recent works \cite{Bhattacharya:2016jce, Benzke:2017yjn,Zenaiev:2019ktw,Jeong:2021vqp,Bai:2022xad}. The dominant contribution to this flux involves Z-moments over production cross sections for charmed hadrons. In the collinear factorisation framework, these cross sections depend on partonic charm production cross sections, which can be computed in perturbative QCD, and the non-perturbative parton distribution functions (PDF) and fragmentation functions (FF) which in turn depend on the factorization scale. The partonic cross sections depend on the charm quark (pole) mass and the factorisation and renormalisation scales. The idea then is to fit the non-perturbative components of PDFs and FFs to as many accelerator-based measurements as possible to minimise uncertainties when extrapolating to energies and phase space inaccessible to direct experiments but relevant for cosmic ray physics.

The resulting uncertainties of prompt neutrino fluxes are relatively large, up to a factor $\simeq 5$ \cite{Zenaiev:2019ktw,Jeong:2021vqp,Bai:2022xad}. At neutrino energies below $\sim 1\,$PeV in the lab frame, they are dominated by QCD uncertainties, where renormalisation and factorisation scale uncertainties play the largest role. PDF uncertainties increase with neutrino energy due to increasing sensitivity to the gluon PDF in the target at very small momentum fraction, where the gluon density rises rapidly and may experience saturation. Above $\sim 1\,$PeV, uncertainties in the all-nucleon flux progressively become comparable to the QCD uncertainties. The former derive to a large degree from theoretical uncertainties in the elemental composition of cosmic rays, and to some extent from inconsistencies between air shower measurements.

The composition uncertainties are dominated by theoretical uncertainties in the event generators that are used to interpret air shower data. Therefore, the accuracy of these calculations above 1\,PeV will indirectly profit from global tuning of event generators. However, one can also turn this argument around and exploit this sensitivity for tuning, by using the atmospheric muon flux as input, which in contrast to the neutrino flux is purely of atmospheric origin and can be measured by neutrino observatories like IceCube.

\section{Details on experiments }
\label{app:exp}

\subsection{Accelerator: ALICE}
ALICE \cite{ALICE:2008ngc} is a general purpose detector for QCD and heavy-ion collision studies at the LHC. It was designed to study strongly interacting matter under high temperature and energy density to investigate the properties of the so-called Quark-Gluon Plasma (QGP). The ALICE detector consists of two main parts. The first component hinges on a central tracking and particle identification system, which exploits energy-loss, transition-radiation, time-of-flight, calorimeter information and more, in order to be able to track and identify hadrons. This system covers the range $|\eta| < 0.9$ and transverse momentum from 0.1 to tens of GeV$/c$. The tracker of ALICE was designed to handle collisions with thousands of particles. In addition to this mid-rapidity region, it has a forward single-arm muon spectrometer covering $-4.0 < \eta < -2.5$, and particle counters that are used to trigger the detector and to measure event activities of the collision (centrality in heavy-ion collisions), covering $-3.7 < \eta < -1.7$, and $2.8 < \eta < 5.1$, respectively. A system of zero-degree calorimeters measures protons and neutrons scattered at small angles (with $|\eta| > 7.0$ typically).

ALICE in its configuration of LHC run 1 [2009-2013] and run 2 [2015-2018] has already provided unique measurements for the study of QCD and the tuning of event generators at the LHC, thanks to i) its sensitivity down to almost zero transverse momentum, where non-perturbative effects dominate, ii) its coverage of \ppColl, \pPbColl, and \PbPbColl collisions, and iii) the particle identification capabilities. A wealth of these results has recently been summarised in Ref.~\cite{ALICE:2022wpn}. Important for this context are the production cross sections for charged particles up to $\eta \sim 5$, and measurements of the hadron composition in multiple systems and as a function of charged-particle multiplicity. In particular, the enhancement of strangeness production counts among the relevant traits to explain the muon puzzle in air showers.

ALICE discovered multiplicity-dependent strangeness enhancement in \ppColl and \pPbColl collisions, where it was not expected \cite{ALICE:2016fzo}. On the one hand, these measurements showed that production and hadronisation in dense final states is modified, and that the classic assumption of universal fragmentation breaks down. On the other hand, this modification was found to follow at first order the charged-particle multiplicity of the event, \ie  to be largely independent of the collision system, so that a modified form of universality still holds. Not only strangeness is affected, hadron composition generally changes in high-multiplicity events. In the light flavour sector, the productions of stable baryons (protons) as well as short-lived resonances (with lifetime below 10 fm/$c$, like $\rho$ and K(892)$^0$) are reduced; the formation of light nuclei (d,t,$^3$He) is enhanced. For these three aspects, this strongly suggests that there exist interactions in the late hadronic stage of the collision, with a significant impact on the collision outcome. In the heavy-flavour sector, \emph{open} charm production (D$^0$, $\Lambda_c^+$, ...) is enhanced as well, but here the effect appears as more dependent on the collision system. Despite these effects, the relative composition of light particles ($\pi$, $K$, $p$) remains fairly constant at mid-rapidity in average events. The strangeness enhancement for the lightest hadrons with one strange quark (K$^0_s$, $\Lambda$) is mild.

Interestingly for our concerns here, ALICE casted recently a family of studies correlating the energy flux in the ZDC in the very forward region versus the event activity in the mid-rapidity region; the novelty is that detailed studies are carried out in \ppColl and \pPbColl systems.
The entry point is a study performed for leading energy in the very forward region versus the multiplicity of unidentified charged particles in the mid-rapidity vicinity \cite{ALICE:2021poe}; the investigations are further developed versus the strange baryons present at mid-rapidity, as most sensitive particles to strangeness enhancement ($\Lambda$, $\Xi$, $\Omega$)\cite{ALICE:2024edc}.
At fixed multiplicity in the mid-rapidity region, strangeness enhancement shows a prevailing correlation with effective energy available outside the ZDC. It thus reveals the strong influence of the initial stage of the collision, , to be compared with the impact of the final-state situation.

\subsection{Accelerator: LHCb}

The LHCb detector \cite{LHCb:2014set} is a single-arm general-purpose forward spectrometer at the LHC covering the range $2 < \eta < 5$ and transverse momentum from 0.1 to tens of GeV$/c$. It was designed to study the decays of hadrons containing heavy quarks produced in high-energy proton-proton collisions. The detector consists of a high-precision tracking system for charged particles with a high-resolution vertex detector very close to the interaction point to reconstruct decay chains of short-lived hadrons, and several sub-detectors that provide particle identification. These consist of two ring-imaging Cherenkov detectors to separate charged pions, kaons, and protons, a calorimeter system for electron/photon and electron/pion separation, and a muon system. The electromagnetic calorimeter is finely segmented with good energy resolution to allow the study of photons and neutral pion decays. It is the only detector at the LHC with these identification capabilities over its entire acceptance.

LHCb provides unique input for tuning event generators through precision measurements in the forward region. Its measurements of $D$ and $B$ meson production constrain parton distribution functions (PDFs) in the proton and in lead nuclei: LHCb is uniquely able to probe the gluon PDFs down to a momentum fraction of $\sim 10^{-6}$, which provides strong constraints for simulations of the atmospheric lepton flux observed by neutrino observatories \cite{Fedynitch:2018cbl,Zenaiev:2019ktw}. In regard to soft-QCD, the production cross section for prompt long-lived charged particles has been recently measured in \ppColl and \pPbColl collisions \cite{LHCb:2021vww,LHCb:2021abm} to an accuracy of a few percent, and neutral pion production has been measured in \ppColl and \pPbColl collisions \cite{LHCb:2022tjh}. Ratios of pions, kaons, and protons have been studied in \ppColl collisions up to 7\,TeV \cite{LHCb:2012lfk}, and an ongoing analysis studies these ratios in \ppColl at 13\,TeV and in \pPbColl at 8.16\,TeV. Following the discovery of multiplicity-dependent strangeness enhancement in \ppColl collisions by ALICE, LHCb found evidence for a multiplicity-dependent rise in the $B_s^0/B^0$ ratio \cite{LHCb:2022syj}, and observed a rise in the $D_s^+/D^+$ \cite{LHCb:2023rpm} ratio.

The LHCb experiment is in the unique position that it can take data in both collider mode and fixed-target mode. This is possible due to LHCb's fixed-target system called SMOG (System for Measuring Overlap with Gas) \cite{Barschel:1693671}. The SMOG system was originally designed for precise luminosity calibration \cite{FerroLuzzi:2005em} of colliding proton beams. By injecting small amounts of noble gases directly into the primary LHC vacuum around the vertex detector, interactions of the LHC beams with gas can be studied in collisions with center-of-mass energies up to 113\,GeV, the highest achieved so far in a fixed-target experiment. This was used to measure the anti-proton production cross section in proton-helium collisions \cite{LHCb:2018ygc,LHCb:2022bbb}, and charm production in proton-neon and lead-neon collisions \cite{LHCb:2022qvj,LHCb:2022sxs}, which constrain a potential intrinsic charm component in the proton with implications for the prompt atmospheric lepton flux.

In recent years, the LHCb detector underwent a major upgrade in preparation for the current data taking period \cite{LHCb:2023hlw}, including a new SMOG2 system \cite{CERN-LHCC-2019-005}. Gas is now injected into an open storage cell located upstream of the LHCb collision point, and suppressed outside it by vacuum pumps. The new system allows one to study collisions with non-noble gases as targets, including hydrogen and deuterium, and possibly oxygen and nitrogen, and at higher gas pressures, increasing the luminosity by two orders of magnitude compared to before \cite{Bursche:2649878, CERN-LHCC-2019-005}. Especially proton-oxygen and proton-nitrogen collisions will be of great interest for air showers. The increased luminosity will also make precision measurements of $D$ and $B$ meson production possible. During the planned run with oxygen beam in 2025, collisions on a hydrogen target will give access to proton-oxygen collisions in the forward hemisphere.

\subsection{Accelerator: LHCf}
LHC forward (LHCf) is a forward experiment at LHC \cite{LHCf:2008lfy} designed to study production of energetic neutral particles, such as $\gamma$, $\piz$ and neutrons, emitted in the very forward region at pseudorapidity $\eta > 8.4$, which significantly contribute to the air shower development induced by high-energy cosmic rays.
The LHCf experiment has two independent subdetectors, called Arm1 and Arm2, which were installed in the instrumental slots of the TANs (Neutral Target Absorbers) located at a distance of $\pm140$\,m from the ATLAS interaction point. Each LHCf detector is composed of two sampling and position-sensitive calorimeter towers. Above the design threshold of 100\,GeV, the energy resolution is better than 5\% for photons and 40\% for neutrons.

Since the LHC start, the LHCf experiment has carried out a number of operations at various collision energies from 0.9 to 13.6 TeV with \ppColl and \pPbColl. The measured inclusive differential cross sections of $\gamma$, $\piz$ and neutron~\cite{LHCf:2012stt,LHCf:2015nel,LHCf:2015rcj,LHCf:2017fnw,LHCf:2018gbv,LHCf:2020hjf}, mainly originating from the fragment of protons, were used to test and tune the hadronic interaction models. The most recent operation in 2022 with \ppColl collisions at $\sqrt{s} = 13.6$\,TeV focused on an increase of statistics to study strange hadron production using measurements of $\eta$  $(\eta \rightarrow 2\gamma)$, $K^0_s$ $(K^0_s \rightarrow 2\pi^0 \rightarrow 4\gamma)$, and $\Lambda$ $(\Lambda \rightarrow n+\pi \rightarrow n+2\gamma)$, which is important to solve the muon puzzle of high-energy cosmic-ray observations, since strangeness enhancement in the forward region would increase the muon yield in air shower simulations.

Furthermore, details of forward particle production such as diffractive processes are studied in joint analyses with the ATLAS experiment. Requiring no particle detection in the ATLAS inner tracker covering the pseudo-rapidity range of $|\eta|\,<\,2.5$, LHCf events originating from low-mass diffractive collisions are easily selected~\cite{ATLAS:2017rme}. In the 2022 operation, a common operation with ATLAS including the ATLAS zero-degree calorimeter (ZDC) and roman-pot detectors (AFP and ALFA) was also performed. Combining data of these ATLAS forward detectors, various physics processes, $p-\pi$ interaction via one-pion-exchange process and low-mass resonance production~\cite{ATLAS:2023ape}, can be studied. These results will be important inputs to tune hadronic interaction models.

\subsection{Accelerator: NA61/SHINE}

NA61/\-SHINE (SPS Heavy Ion and Neutrino Experiment)~\cite{Abgrall:2014fa} is a multipurpose fixed-target experiment designed to study hadron production in hadron-nucleus and nucleus-nucleus collisions at the CERN Super Proton Synchrotron (SPS). The core component of the detector comprises a set of large-acceptance Time Projection Chambers (TPCs) and two superconducting magnets with a combined bending power of 9\,Tm. This setup enables precise measurement of particle momenta ($\sigma(p)/p^2\approx (0.3{-}7){\times}10^{-4}\,\text{GeV}/c$) and provides excellent particle identification capabilities via the specific energy loss in the TPC volumes.

The experiment commenced operations in 2007 and has since gathered hadroproduction data using a variety of projectiles, beam energies, and target materials. NA61/SHINE conducted detailed measurements of particle production in p+C interactions at 31 GeV/c~\cite{Abgrall:2011ae, Abgrall:2011ts, Abgrall:2015hmv} and \pipCColl at 60 GeV/c to determine the beam properties in accelerator-based neutrino experiments~\cite{NA61SHINE:2019nzr}. Since carbon is a good proxy for interactions on nitrogen in air, the measured spectra in these reactions are also highly relevant for modelling low-energy interactions in air showers. Of particular interest for air shower physics are the measurements of particle production in interactions of negatively charged pions with carbon at 158 and 350 GeV/c~\cite{NA61SHINE:2017vqs,NA61SHINE:2022tiz}. The spectra of $\pi^\pm$, K$^\pm$, p, $\bar{p}$, $\rho^0$, $\omega$, K$^{*0}$, K$^{0}_\text{S}$, $\Lambda$, $\bar{\Lambda}$ were measured in a wide range of longitudinal and transverse momentum.

Furthermore, p+p interactions were studied in a wide range of beam momenta (20, 31, 40, 80 and 158~GeV/c)~\cite{Abgrall:2013pp_pim, Aduszkiewicz:2017sei} to establish a reference data set for the heavy ion physics program of NA61/SHINE. This data set is also instrumental in studying the secondary production of anti-protons and anti-nuclei in collisions of cosmic-ray protons with the interstellar medium in the Galaxy.

Finally, following a successful pilot run in 2018~\cite{Unger:2019nus,Amin:2023fki}, the collaboration plans to take a substantial data set on the fragmentation cross sections of nuclei. These data are essential to interpret the recent high-precision cosmic-ray data on secondary Galactic nuclei, see \eg Ref.~\cite{Genolini:2018ekk,Genolini:2023kcj} and will also help refine the fragmentation models of hadronic interaction models, thereby improving the modelling of air shower fluctuations.

\subsection{Astroparticle: Pierre Auger Observatory}

The Pierre Auger Observatory~\cite{PierreAuger:2015eyc} is located in Malargüe, Argentina, at an atmospheric depth of $880\,\rm{g/cm}^2$. It is the world's largest observatory for the detection of ultra-high-energy cosmic rays and has been operating successfully for almost 20 years.
The hybrid design of the Observatory consists of a Surface Detector (SD) array of 1660 water Cherenkov tanks arranged in a triangular grid with a baseline of 1.5\,km, covering an area of $3000\,\text{km}^2$. It is complemented by the Fluorescence Detector (FD)~\cite{Abraham:2009pm}, which consists of four telescope sites, each with 6 telescopes, at the periphery of the Observatory overlooking the SD at elevation angles from $0^\circ$ to $30^\circ$. Another set of three telescopes (High Elevation Telescopes; HEAT) extends the elevation range to $60^\circ$ above the horizon. Buried muon detectors made of plastic scintillator slabs provide additional clean information about air-shower muons.

A fraction of the events (hybrid events) are observed in both SD and FD. It has been shown that the combination of signals from surface and fluorescence detectors is a powerful tool to examine current models of hadronic interactions~\cite{PierreAuger:2024neu}.Five two-dimensional distributions, corresponding to different cosmic ray mass hypotheses, of ground signal at 1000\,m and depth of shower maximum ($X_\text{max}$) were simultaneously fitted with Monte-Carlo templates. Surprisingly, the predicted $X_\text{max}$ scale for the models EPOS-LHC, QGSJET-II-04 and \sibyll{}-2.3d describes Auger data better assuming a depth in the atmosphere which is deeper by about 20 to 50\,g/cm$^{2}$, obtaining a mixed mass composition of primary particles that is heavier than for unmodified $X_\text{max}$ scales of the models. The predicted hadronic signal should be increased by about 15-25\%, which alleviates the muon puzzle for unmodified $X_\text{max}$ scales. The significance of improvement in the description of measured data with the assumed modifications of model predictions has been found to be more than 5$\sigma$ for any linear combination of experimental systematic uncertainties. 
Interestingly, from the zenith-angle dependence of the fitted hadronic scale, there is a strong indication ($4.4\sigma$) of too hard energy spectra of muons predicted by QGSJET-II-04. In Ref.~\cite{Ulrich:2010rg} the impact of modifications of basic parameters of hadronic interactions has been studied in detail using the 1-D simulation package \conex{}, in particular, multiplicity, elasticity, and cross section. 
It is possible to apply the same approach to study general 3-D observables, applying different ranges and thresholds to the modifications of individual parameters based on existing experimental constraints and allow the modification of all three parameters at once~\cite{Ebr:2023nkf, Blazek:2021jzf, Blazek:2023bbg}. Due to the general anti-correlation between the change of the muon signal at 1000\,m, $R_\mu(1000)$, and $X_\mathrm{max}$, only a very specific combination of maximal considered modifications (decreased cross section, and increased elasticity and multiplicity) fits the recent Auger results in the $R_\mu(1000)$--$X_\mathrm{max}$ plane~\cite{PierreAuger:2024neu}. Such modifications are, however, in tension with the Auger measurements of the proton-air cross section~\cite{PierreAuger:2012egl} and possibly with the higher moments of the observed $X_\mathrm{max}$ distributions.

Now in its second phase of operation, the Pierre Auger Observatory is undergoing a major upgrade, known as AugerPrime~\cite{thepierreaugercollaboration2016pierre}, to increase its sensitivity to the primary mass. The Underground Muon Detector (UMD) will be installed in the low-energy extension of the surface detector as part of the upgrade. It consists of a \qty{30}{m^2} array of plastic scintillator muon counters buried \qty{2.3}{m} underground near the water-Cherenkov detectors of a nested infill array. UMD will provide a direct measurement of the muon component of air showers in the energy range \qty{3e16}{eV} to \qty{1e19}{eV}. This will contribute significantly to the discrimination of the primary mass and to the testing of hadronic interaction models. The soil above each UMD detector is responsible for absorbing the electromagnetic component of air showers, and imposes an energy cut of about \qty{1}{GeV} for vertical muons. The muon counts are used to fit a lateral distribution function and the resulting muon density at a reference distance of \qty{450}{m} used for comparisons with simulated showers. Preliminary data of a subset of stations equipped with PMTs were used to quantify a muon deficit of the order 35-50\% in comparison to data of corresponding X$_\text{max}$ measurements. Future measurements of the entire infilled area of about \qty{24}{m^2}, exploiting SiPM sensors~\cite{Aab_2017} to replace the former PMTs, will provide a high-statistics measurement of the muon content and will allow calibrating muon estimates of the upgraded surface stations. 

As part of the AugerPrime upgrade, the surface detector stations of the Pierre Auger Observatory were equipped each with a dual-polarized Short Aperiodic Loaded Loop Antenna
(SALLA) measuring the radio signals from EAS in the 30–80MHz band. This extension allows measuring the energy in the electromagnetic cascade of inclined air showers with zenith angles above around $65^{\circ}$ with a resolution of below 10\%~\cite{Huege:2023pfb}. Combining the measurement of this electromagnetic energy with the pure measurement of the muon content of inclined air showers by the water-Cherenkov detectors will enable precise studies of the number of muons and their fluctuations at energies beyond $10^{18.4}$\,eV, with much larger statistics than could be achieved previously through a combination of water-Cherenkov and fluorescence detector data. Furthermore, the radio detector will allow to cross-check the energy scale of cosmic-ray measurements with an independent approach based on the first-principle classical electrodynamics calculation of radio signals from EAS \cite{PierreAuger:2016vya}.

\subsection{Astroparticle: IceCube Neutrino Observatory}

The IceCube Neutrino Observatory~\cite{IceCube:2016zyt} (IceCube) is a cubic-kilometer detector deployed deep in the ice at the geographic South Pole at depths between $1450$\,m and $2450$\,m, accompanied by a surface detector array, IceTop~\cite{2013NIMPA.700..188A}. The in-ice detector measures high-energy muons above a few $100$\,GeV from EAS that penetrate the Antarctic ice, as well as charged secondaries induced by neutrino interactions. In addition, IceTop~\cite{2013NIMPA.700..188A} measures EAS from cosmic-ray interactions from PeV to EeV energies at an atmospheric depth of about $690\,\rm{g/cm}^2$. This hybrid detector setup provides unique opportunities to study hadronic interactions in EAS in great detail. New surface detectors for IceTop~\cite{Haungs:2019ylq} and IceCube-Gen2~\cite{IceCube:2014gqr} are under development, including scintillator detectors and radio antennas, which will further enhance the capability to study hadronic interaction models in the future.

IceTop has recently reported a measurement of the muon densities in EAS as functions of energy at reference distances of 600\,m and 800\,m for primary energies between 2.5\,PeV and 40\,PeV and between 9\,PeV and 120\,PeV, respectively~\cite{IceCubeCollaboration:2022tla}.
These measurements are consistent within uncertainties with predictions using the pre-LHC model \sibyll{}-2.1. However, comparisons to simulations using the post-LHC models \eposlhc{} and \qgsjet{}-II-04 yield higher muon densities than observed. Interestingly, preliminary results of a recent measurement of the multiplicity of TeV muons in EAS, measured in the deep ice with IceCube~\cite{Verpoest:2023qmq}, show agreement within uncertainties with all hadronic interaction models considered. These two measurements therefore indicate inconsistencies in the modelling of GeV and TeV muons within the post-LHC models. To further study these inconsistencies, efforts for a simultaneous measurement of GeV and TeV muons from the same air shower on an event-by-event basis are ongoing~\cite{IceCube:2023lhg,IceCube:2023suf}. These measurements will yield important information about the energy sharing between low-energy and high-energy interactions during the EAS development and provide strong constraints for hadronic interaction models~\cite{Riehn:2019jet}.

IceCube's deep-ice detector also allows for measurements of the atmospheric neutrino spectrum~\cite{IceCube:2023qpn, IceCube:2020acn, IceCube:2023lad, IceCube:2021uhz}, as well as the muon energy spectrum at very high energies (above a few $10$\,TeV)~\cite{IceCube:2015wro,Fuchs:2017nuo,Soldin:2018vak}. The combination of these two channels  provides valuable input to hadronic interaction models, such as the kaon/pion ratio, or the contribution from the prompt atmospheric lepton flux~\cite{Fedynitch:2018cbl}. Current efforts are underway to update these measurements and to utilize their synergies in constraining hadronic interaction models to measure the relative contributions from the decay of unflavoured and charmed mesons, for example. In addition, measurements of the seasonal variations of the high-energy muon~\cite{Tilav:2019xmf,IceCube:2021eth} and neutrino fluxes~\cite{IceCube:2023qem} also yield information on the kaon/pion ratio~\cite{Desiati:2010wt,Verpoest:2024dmc} to further constrain hadronic interaction models.

\subsection{Astroparticle: KASCADE}

The cosmic-ray air-shower experiments KASCADE~\cite{KASCADE:2003swk} and KASCADE-Grande~\cite{Apel:2010zz} were located at the Karlsruhe Institute of Technology, Germany, at an atmospheric depth of $1022\,\rm{g/cm}^2$. 
These detector arrays measured the energy spectrum and mass composition of cosmic rays in the primary energy range of PeV to EeV~\cite{KASCADE:2005ynk,KASCADEGrande:2011kpw,Apel:2013ura}. 
The results have been compared to predictions from different hadronic interaction models~\cite{KASCADE-Grande:2022dog}. In general, agreement between the models for the all-particle cosmic ray flux are better than for its composition. Interestingly, post-LHC models show a lower all-particle flux than the pre-LHC models at energies of around PeV.

The special configuration of the experiments allows one to combine the KASCADE and KASCADE-Grande data with the possibility to investigate and compare the muon content close to the shower core with those farther away of the core~\cite{KASCADE-Grande:2017ddf}. A combined analysis shows a systematic deficiency for the hadronic interaction models \qgsjet{}-II-04, \eposlhc{}, and
\sibyll{}-2.3 to describe the muon content of the showers consistently. Either there are too few muons predicted in the center of the shower or too few at larger distances. This is true for all models considered in the analysis.

In addition, an analysis was performed to estimate the muon content in cosmic-ray induced air showers as a function of the primary energy~\cite{KASCADEGrande:2021lvu}. These measurements were compared to predictions of the event generators \qgsjet{}-II-04, \eposlhc{}, and \sibyll{}-2.3, which were not able to consistently describe the KASCADE-Grande data for all zenith angles and energies. This indicates that the attenuation of the number of muons with the zenith angle is smaller in data than in simulations. The observed anomalies could imply that the energy spectrum of muons from real EAS at production site for a given primary energy is harder than the respective model predictions. Further investigations are ongoing, even after the completion of the measurements, using data that is available through an open portal, the KASCADE Cosmic Ray Data Centre (KCDC)~\cite{Haungs:2018xpw}.

\subsection{The muon puzzle in EAS}
\label{app:exp-muon-puzzle}

To further investigate the muon deficit reported by Auger and other experiments, the Working Group for Hadronic Interactions and Shower Physics (WHISP)~\cite{EAS-MSU:2019kmv,Cazon:2020zhx,Soldin:2021wyv,ArteagaVelazquez:2023fda} has performed a meta-analysis of muon number measurements from different air-shower observatories.
This group is formed by members of the EAS-MSU, IceCube, KASCADE-Grande, NEVOD-DECOR, Pierre Auger, SUGAR, Telescope Array (TA), and Yakutsk EAS Array collaborations. The group combined published data from a diverse set of muon detection methods including ice-Cherenkov stations of IceTop~\cite{IceCubeCollaboration:2022tla}, shielded scintillating detectors of KASCADE-Grande~\cite{KASCADE-Grande:2017wfe}, underground scintillation detectors of the Yakutsk array~\cite{Glushkov:2023ani}, the underground Geiger-Mueller counters of EAS-MSU~\cite{Fomin:2016kul}, the tracking detector and water-Cherenkov calorimeter of NEVOD-DECOR~\cite{Bogdanov:2010zz,Bogdanov:2018sfw}, the underground liquid-scintillator tanks of SUGAR~\cite{Bellido:2018toz,Kalmykov:2022grb}, the buried scintillator counters of HiRes-MIA~\cite{HiRes:1999ioa}, the scintillator modules of TA~\cite{TelescopeArray:2018eph}, the water-Cherenkov array of Auger~\cite{PierreAuger:2014ucz,PierreAuger:2016nfk,PierreAuger:2021qsd}, and the underground scintillator modules of Auger~\cite{PierreAuger:2020gxz}, as well as the shielded scintillator array of AGASA~\cite{Gesualdi:2021yay}, alongside previously unconsidered data from Haverah Park~\cite{Cazon:2023ojj} and new estimates from an analysis of KASCADE-Grande data~\cite{ArteagaVelazquez:2023nqa} that uses the energy scale of the Pierre Auger Observatory for calibration.
The diversity of these measurements prevents a direct comparison. Apart from the cosmic ray energy $E$, the observed muon density at ground level depends on the atmospheric depth of the ground array, the lateral distance at which the muon density is measured, the zenith angle of the showers considered, and the effective energy cutoff introduced by shielding of detectors.
Furthermore, the muon number $N_{\mu}$ is measured in many different ways. The standard way is to use shielded detectors to isolate the muon signal from other particles produced in the air shower, but experiments without such shielded detectors have used other techniques; some used the atmosphere itself as a shield by analysing highly inclined showers, or discriminate muon hits in ground detectors from hits of other particles based on the characteristic energy deposit of muons, which peaks around a value that depends on the muon inclination.
The comparison of the experimental data and the model expectations was therefore done using the $z$ scale, which compares the respective measurements with the expectation from simulated air showers. It is defined as
\begin{equation}
    \label{eq:z-scale}
    z=\frac{\ln\langle N_\mu^{\rm{det}} \rangle(E) - \ln\langle N_{\mu,\rm{p}}^{\rm{det}} \rangle(E)}{\ln\langle N_{\mu,\rm{Fe}}^{\rm{det}} \rangle(E) - \ln\langle N_{\mu,\rm{p}}^{\rm{det}} \rangle(E)}
\end{equation}\\
where $\langle N_\mu^\mathrm{det}\rangle$ is the mean value of the measured muon density under the specific conditions of the experiment, and $\langle N_{\mu,\mathrm{p}}^\mathrm{det}\rangle$  ($\langle N_{\mu,\mathrm{Fe}}^\mathrm{det}\rangle$ ) is the corresponding prediction of the average muon density for proton (iron) showers under the same conditions.

The method of estimating the shower energy $E$ is important for the $z$-scale, since $\nmu$ is nearly proportional to $E$. Systematic shifts in the energy scale lead to apparent shifts in the $z$-values. To address this, the WHISP adjusted the energy scales of the experiments to align them with a common reference spectrum at a given energy \cite{EAS-MSU:2019kmv}. This greatly improved the consistency of the measurements.

Furthermore, correlations between the estimation of the muon density $\nmu$ and the energy $E$ can distort the measurement, and should ideally be negligible.
The cosmic ray energy $E$ is ideally estimated by integrating the longitudinal energy loss profile in the atmosphere observed with fluorescence or Cherenkov telescopes. Experiments without telescopes instead use the charged particle density measured in surface detector arrays, which includes a contribution from muons. In the latter case, the energy estimate always shows some degree of cross-contamination from $N_{\mu}$, introducing positive correlations event-by-event. Finally, some experiments measure only the muon number $\nmu$ and obtain a muon multiplicity distribution. This distribution can be compared with predictions based on the cosmic ray flux from another experiment obtained with the FD technique to infer $z$-values (sometimes referred to as \emph{intensity based} $z$-values).

In addition to an uncorrelated energy estimation method -- ideal for a muon measurement with small systematic uncertainties -- a full detector simulation is required to account for potential detector biases and a small atmospheric overburden. The large atmospheric overburden is irrelevant for an FD-based energy estimation, but increases the uncertainties for a ground-based estimation.

\begin{figure}[tb]
    \centering
    \includegraphics[width=\textwidth]{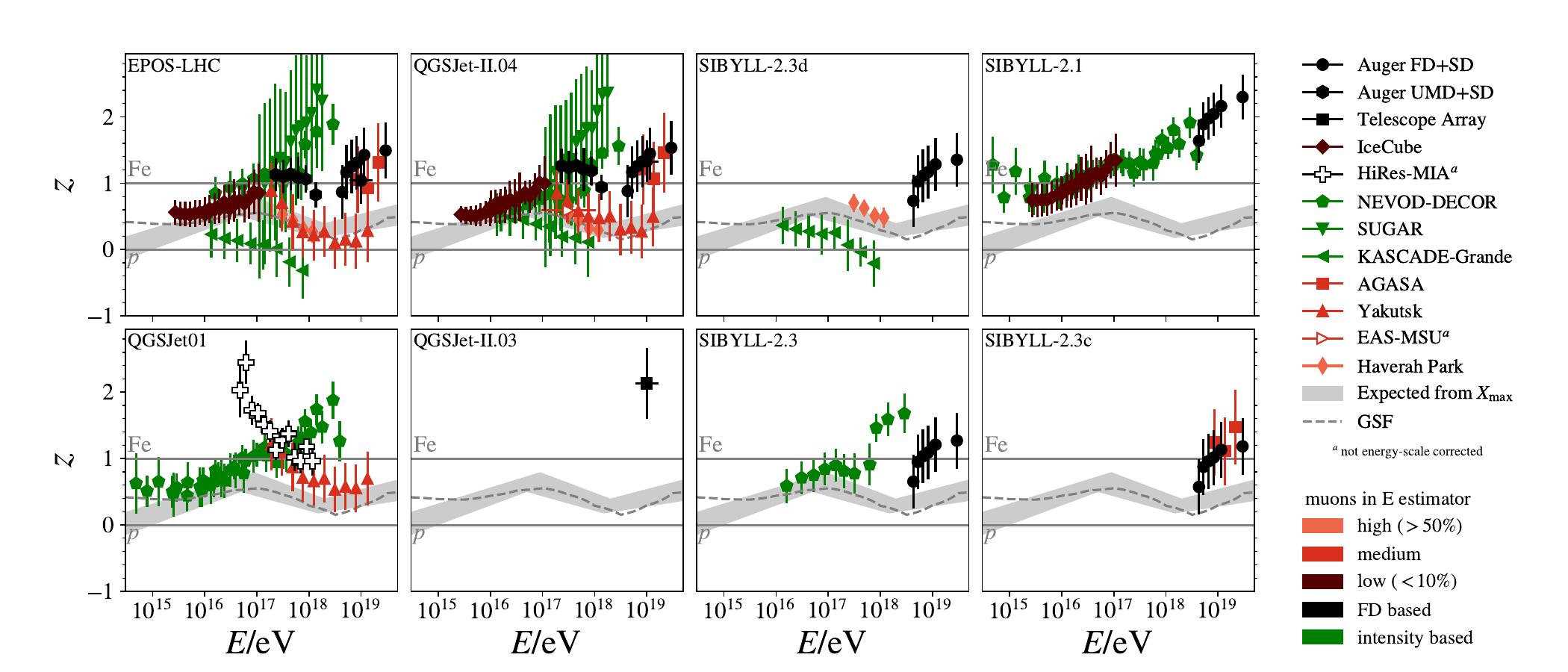}
    \caption{Muon content of air showers encoded in z-values (see text) as function of shower energy
E from different experiments. The event generator used to compute the predicted muon content is
shown in the upper left corner of each plot. The colors indicate how much muons contribute to the
estimate of the shower energy E. The dashed line indicated the expected z-value based on the GSF
model~\cite{Dembinski:2017zsh}, while the gray band shows the expectation from Auger \xmax measurements.  Error bars show statistical and systematic uncertainties added in quadrature. Figure taken from Ref.~\cite{ArteagaVelazquez:2023fda}}.
    \label{fig:WHISP-all}
\end{figure}

The results of the cross-calibrated $z$ values are shown in \autoref{fig:WHISP-all}. The $z$ values depend on the event generator used in the air shower simulations. The WHISP group uses published simulations that employed event generators available at the time of publication. Therefore, not all data points are available for each event generator. 
At first glance, there is no globally coherent picture from 1 PeV to 10 EeV. Taking into account the different conditions under which the experiments performed the measurements, such as distance to the shower core, zenith angle, or muon energy, does not reduce the problem~\cite{Cazon:2020zhx}. However, two groups can be identified. Experiments that directly measure the shower energy with little (red brown markers) or no muon contribution (black markers), such as IceCube and Auger, show a muon deficit in the simulations that grows at a constant rate with increasing energy and appears as an increase in $z$ over the expectation from the cosmic ray composition. Experiments with medium (red) or high contribution (orange) of muons to the estimator of the shower energy, or intensity based experiments (green) where the energy scale is inferred from the muon content itself, show no consistent picture. Presumably because the strong dependence of the number of muons on the shower energy masks the deficit.

\section{Details on tuning}
\label{app:tuning}

\begin{table}[tbp]
    \centering
    \tablecaption{\rivet plugins and HEPData entries used in the forward tuning of Pythia, along with the ongoing \rivet-ization efforts toward a tune for air showers, as discussed in \autoref{sec:astro-tune}. The third column specifies whether the analysis has an entry in the HEPData database, while the fourth column indicates whether the plugin is included in the public \rivet repository. The fifth and sixth columns define the collision system, where $\sqrt{s}$ represents the center-of-mass energy for proton-proton collisions, and $\sqrt{s_\text{NN}}$ denotes the energy in the nucleon-nucleon frame; the second half of the table focuses on fixed-target experiments, organized by beam momentum. The final column lists the extracted final states used for tuning.}
    \label{tab:rivet_table}
    \small
    \hspace*{-1cm}\begin{tabular}{l c c c c c c}
        \toprule
        \rivet plugin                 & Ref.                                  & HEPData                                                        & Published
                                      & $\sqrt{s}$, $\sqrt{s_\text{NN}}$ (TeV) & Collision system  & Final state / Observable                                                                                                                                                            \\
                                      \midrule

        \texttt{LHCF\_2015\_I1351909} & \cite{LHCf:2015nel}                    & \href{https://www.hepdata.net/record/ins1351909}{$\checkmark$} & \href{https://rivet.hepforge.org/analyses/LHCF_2015_I1351909.html}{$\checkmark$}
                                      & 7                                      & \TppColl    & neutral $\mathrm{d}\sigma/\mathrm{d}x_{\rm F}$                                                                                                                                                                   \\
        \texttt{LHCF\_2016\_I1385877} & \cite{LHCf:2015rcj}                    & \href{https://www.hepdata.net/record/ins1385877}{$\checkmark$} & \href{https://rivet.hepforge.org/analyses/LHCF_2016_I1385877.html}{$\checkmark$}
                                      & 2.76, 7, 5.02                          & \TppColl, \TpPbColl & neutral $\mathrm{d}\sigma/\mathrm{d}x_{\rm F}$                                                                                                                                                            \\
        \texttt{LHCF\_2018\_I1692008} & \cite{LHCf:2020hjf}                    & \href{https://www.hepdata.net/record/ins1783943}{$\checkmark$} & \href{https://rivet.hepforge.org/analyses/LHCF_2018_I1692008.html}{$\checkmark$}
                                      & 13                                     & \TppColl           
                                      & neutral $\mathrm{d}\sigma/\mathrm{d}x_{\rm F}$ \\
        \texttt{LHCF\_2018\_I1518782} & \cite{LHCf:2017fnw}                    & \href{https://www.hepdata.net/record/ins1518782}{$\checkmark$} & \href{https://rivet.hepforge.org/analyses/LHCF_2018_I1518782.html}{$\checkmark$}
                                      & 13                                     & \TppColl           
                                      & neutral $\mathrm{d}\sigma/\mathrm{d}x_{\rm F}$ \\
        \texttt{LHCF\_2023\_I2658888} & \cite{Piparo:2023yam}                  & \href{https://www.hepdata.net/record/ins2658888}{$\checkmark$} & \href{https://rivet.hepforge.org/analyses/LHCF_2023_I2658888}{$\checkmark$}
                                      & 13                                     & \TppColl           
                                      & neutral $\mathrm{d}\sigma/\mathrm{d}x_{\rm F}$\\
        \midrule
        \texttt{LHCB\_2013\_I1251899} & \cite{LHCb:2013gmv}                     & \href{https://www.hepdata.net/record/ins1251899}{$\checkmark$} & $\checkmark$                                                                                                                        & 5                                     & \TpPbColl & $\mathrm{d}\sigma/\mathrm{d}p_{\rm T}\mathrm{d}y$\\
        \texttt{LHCB\_2016\_I1504058} & \cite{LHCb:2016qpe}                     & \href{https://www.hepdata.net/record/ins1504058}{$\checkmark$} & \href{https://rivet.hepforge.org/analyses/LHCB_2016_I1504058.html}{$\checkmark$}                                                                                                                        & 7, 13                                  & \TppColl & $\mathrm{d}\sigma/\mathrm{d}\eta$ \\
        \texttt{LHCB\_2019\_I1720413} & \cite{LHCb:2019avm}                     &  \href{https://www.hepdata.net/record/ins1720413}{$\checkmark$}        & \href{https://rivet.hepforge.org/analyses/LHCB_2019_I1720413.html}{$\checkmark$}                                                                                                                        & 8.16                                   & \TpPbColl & $\mathrm{d}\sigma/\mathrm{d}p_{\rm T}\mathrm{d}y$ \\
        \texttt{LHCB\_2021\_I1889335} & \cite{LHCb:2021abm}                     & \href{https://www.hepdata.net/record/136099}{$\checkmark$}            & \href{https://rivet.hepforge.org/analyses/LHCB_2021_I1889335}{$\checkmark$}                                             & 13                                     & \TppColl & charged $\mathrm{d}\sigma/\mathrm{d}p_{\rm T}\mathrm{d}\eta$ \\
        \texttt{LHCB\_2021\_I1913240} & \cite{LHCb:2021vww}                     & \href{https://www.hepdata.net/record/ins1913240}{$\checkmark$} & \href{https://rivet.hepforge.org/analyses/LHCB_2021_I1913240.html}{$\checkmark$}                                                                                                                        & 5                                      & \TppColl, \TpPbColl & charged $\mathrm{d}\sigma/\mathrm{d}p_{\rm T}\mathrm{d}\eta$ \\
        \texttt{LHCB\_2022\_I2694685} & \cite{LHCb:2022dmh}                     &  \href{https://www.hepdata.net/record/ins2694685}{$\checkmark$}       & \href{https://rivet.hepforge.org/analyses/LHCB_2022_I2694685.html}{$\checkmark$}                                                                                                                         & 8.16                                   & \TpPbColl & $\mathrm{d}\sigma/\mathrm{d}p_{\rm T}\mathrm{d}y$ \\
        
        \midrule
                                      &                                        &                                                                &                                                                                  & $p_{\rm{beam}}$ (GeV/c) & \\
        \midrule
        \texttt{E104\_1976\_I98502}   & \cite{Carroll:1975xf}                  & \href{https://www.hepdata.net/record/ins98502}{$\checkmark$}   &
                                      & [23 - 280]                             & \TpipmpColl, \TKpmpColl, \TppColl, \TppbarColl   
                                      & $\sigma^{\rm tot}$ \\
        \texttt{E104\_1979\_I132765}  & \cite{Carroll:1978vq}                  & \href{https://www.hepdata.net/record/ins132765}{$\checkmark$}  &
                                      & [200 - 370]                            & \TpipmpColl, \TKpmpColl, \TppColl, \TppbarColl                               
                                      & $\sigma^{\rm tot}$\\
        \texttt{E104\_1979\_I132133}  & \cite{Carroll:1978hc}                  & \href{https://inspirehep.net/literature/132133}{$\checkmark$}  &
                                      & [60 - 280]                             & \TpipmpColl, \TKpmpColl, \TppColl, \TppbarColl                               
                                      & $\sigma^{\rm absorption}$\\
        \texttt{NA8\_1983\_I182455}   & \cite{Burq:1982ja}                     & \href{https://www.hepdata.net/record/ins182455}{$\checkmark$}  &
                                      & [30 - 345]                             & \TpimpColl, \TppColl                                               
                                      & elastic $\mathrm{d}\sigma/\mathrm{d}t$ \\
        \texttt{NA22\_1986\_I18431}  & \cite{NA22:1986nhr}                    & \href{https://www.hepdata.net/record/ins18431}{$\checkmark$}  & \href{https://rivet.hepforge.org/analyses/NA22_1986_I18431}{$\checkmark$}
                                      & 250                                    & \TpippColl, \TKppColl, \TppColl                                           
                                      & $P_n$\\
        \texttt{NA22\_1987\_I246909}  & \cite{NA22:1987lmr}                    & \href{https://www.hepdata.net/record/ins246909}{$\checkmark$}  &
                                      & 250                                    & \TpippColl, \TKppColl                                           
                                      & elastic $\mathrm{d}\sigma/\mathrm{d}t$\\
        \texttt{NA22\_1988\_I265504}  & \cite{EHSNA22:1988fqa}                 & \href{https://www.hepdata.net/record/ins265504}{$\checkmark$}  & \href{https://rivet.hepforge.org/analyses/EHS_1988_I265504}{$\checkmark$}
                                      & 250                                    & \TpippColl, \TKppColl, \TppColl                                            
                                      & $\mathrm{d}\sigma/\mathrm{d}x_{\rm F}\mathrm{d}p_{\rm T} \mathrm{d}y$\\
        \texttt{NA22\_1990\_I301243}  & \cite{EHSNA22:1990vem}                 & \href{https://www.hepdata.net/record/ins301243}{$\checkmark$}  &
                                      & 250                                    & \TpippColl         
                                      & $\mathrm{d}\sigma/\mathrm{d}x_{\rm F}, \mathrm{d}\sigma/\mathrm{d}p_{\rm T}^2$\\
        \texttt{NA22\_1992\_I322980}  & \cite{EHSNA22:1991dhh}                 & \href{https://www.hepdata.net/record/ins322980}{$\checkmark$}  &
                                      & 250                                    & \TpippColl, \TKppColl 
                                      & $\mathrm{d}\sigma/\mathrm{d}x_{\rm F}, \mathrm{d}\sigma/\mathrm{d}p_{\rm T}^2$\\
        \texttt{NA49\_2006\_I694016}  & \cite{NA49:2005qor}                    & \href{https://www.hepdata.net/record/ins694016}{$\checkmark$}  & \href{https://rivet.hepforge.org/analyses/NA49_2006_I694016.html}{$\checkmark$}
                                      & 158                                    & \TppColl           
                                      & $E~ \mathrm{d}\sigma/\mathrm{d}x_{\rm F} \mathrm{d}p_{\rm T}^2$\\
        \texttt{NA49\_2009\_I818217}  & \cite{NA49:2009brx}                    &                                                                & \href{https://rivet.hepforge.org/analyses/NA49_2009_I818217.html}{$\checkmark$}
                                      & 158                                    & \TppColl           
                                      & $E~ \mathrm{d}\sigma/\mathrm{d}x_{\rm F} \mathrm{d}p_{\rm T}^2$\\ 
        \texttt{NA61\_2017\_I1598505} & \cite{NA61SHINE:2017fne}               & \href{https://www.hepdata.net/record/ins1598505}{$\checkmark$} &
                                      & [20 - 158]                             & \TppColl           
                                      & $\mathrm{d}^2n/\mathrm{d}y \mathrm{d}p_{\rm T}$\\
        \texttt{NA61\_2017\_I1600971} & \cite{NA61SHINE:2017vqs}               &                                                           &
                                      & 158, 350                               & \TpimCColl         
                                      & $x_{\rm F}~ \mathrm{d}n/\mathrm{d}x_{\rm F} $  \\
        \texttt{NA61\_2019\_I1753094} & \cite{NA61SHINE:2019aip}               & \href{https://www.hepdata.net/record/ins1753094}{$\checkmark$} &
                                      & 60, 120                                & \TpCColl, \TpBeColl, \TpAlColl                                           
                                      & $\sigma^\mathrm{prod}, \sigma^\mathrm{inel}$\\
        \texttt{NA61\_2022\_I1868367} & \cite{NA61SHINE:2021iay}               &                                                                &
                                      & 158                                    & \TppColl         
                                      & $\mathrm{d}^2n/\mathrm{d}y\mathrm{d}p_{\rm T}$, $\mathrm{d}n/\mathrm{d}y$ \\
        \texttt{NA61\_2023\_I2155140} & \cite{NA61SHINE:2022tiz}               &                                                                &
                                      & 158, 350                               & \TpimCColl         
                                      &  $\sigma^\mathrm{prod}$, $\mathrm{d}^2n/\mathrm{d}p\mathrm{d}p_{\rm T}$ \\
        \bottomrule
    \end{tabular}
    \label{fig:tuning}
\end{table}

In automatic tuning, there are two practical approaches commonly used to find an optimal a description of the data: either by minimizing the
least-squares-type cost function via gradient descent, or in a Bayesian framework by computing the
posterior probability density of the parameters from the likelihood function and priors

\paragraph{Gradient-based tuning}
\label{par:gradient-based tuning}

A least-squares-type cost function quantifies the agreement between the event generator and the measurements. It can be optimized based on local gradient descents to determine the best set of tuning parameters. To allow analytical computations of the gradient, the event generator is replaced by a surrogate model.

Gradient-based automatic tuning first became available in particle physics with the \textsc{Professor}~\cite{Buckley:2009bj} software package, which uses a high-dimensional polynomial as a surrogate model. This inspired 
the development of \textsc{Apprentice}~\cite{Krishnamoorthy:2021nwv}, which offers several improvements. It supports the use of Padé approximants (ratios of polynomials) to construct the surrogate model, which provide more flexibility, and provides algorithms to avoid unwanted poles in their construction. 

\paragraph{Bayesian tuning}
Bayesian tuning also starts with a weighted least-squares-type cost function, and the construction of a surrogate model to replace the event generator. A Markov-Chain Monte-Carlo is then used to sample points from the posterior distribution to obtain the best set of parameters, including uncertainties and their correlations. The posterior distribution can be used to obtain the best set of parameters and to study the uncertainties of the parameters and their correlations. The effectiveness of the Bayesian tuning approach was demonstrated using collider data from LEP experiments~\cite{LaCagnina:2023yvi,Schulz:2020ebm}.

An essential first step towards global tuning to particle and astroparticle data is to demonstrate the feasibility of tuning to air shower data. To this end, a study based on the event generator \pythia~8 and the air shower simulation code \corsika~8~\cite{Engel:2018akg} was performed by generating mock air shower data using the default settings in \pythia~8. Bayesian tuning was then used to test whether these default values could be recovered. In one scenario the number of muons produced in the shower, $\nmu$, was chosen as the air shower observable to tune to. In the second scenario, both $\nmu$ and the depth of the shower maximum, $\xmax$ were used. In the first scenario, the strong coupling constant $\alpha_s$ was chosen as the tuning parameter, while in the second scenario, two parameters related to Lund string fragmentation (and thus hadron multiplicity) were chosen. 
Tuning was then performed as described above. A surrogate model was constructed by running multiple sets of air shower simulations of 10\,PeV protons with different values for the respective tuning parameters. In both scenarios, the Bayesian method successfully generated posterior distributions centered on the input values. This study shows that tuning to air shower observables is in principle possible, and that the Bayesian tuning approach is a promising method for global tuning to particle and astroparticle data. The experiment also highlights the need for fast air shower simulations. The next step is to perform a first tune to data.

\vspace{5mm}
\autoref{tab:rivet_table} provides a summary of the \rivet plugins and HEPData entries used in the forward tune of \pythia{}, along with the ongoing efforts towards a tune for air showers discussed in \autoref{sec:astro-tune}.

\subsection{Tuning without a surrogate model} 
\label{app:tuning without surrogate model}

One limitation of established automatic tuning methods is the need to construct a surrogate model. This step adds complexity because the parameter space must be efficiently sampled, and the approach suffers from the curse of dimensionality: to construct a surrogate model, one must perform a grid-like sampling of a high-dimensional parameter space. The number of required grid points grows exponentially with the number of dimensions.

Methods based on stochastic gradient descent (SGD) could potentially be used to tune the event generator directly, avoiding the construction of the surrogate model, and dramatically reducing the computational cost of tuning many parameters at once. SGD algorithms, such as \textsc{Adam} \cite{Kingma:2014vow}, successfully train neural networks with huge parameter spaces at a fraction of the computational cost of traditional methods. They avoid the curse of dimensionality, since the cost of computing the gradient grows only linearly with the number of dimensions.

When training neural networks, gradient formulas are exact, but computed over a random subset of the full data sample. This introduces random fluctuations into the gradient, similar to those introduced by Monte Carlo simulation in the event generator. SGD algorithms are designed to handle these fluctuations, but exploding gradients are still a problem. In the tuning application, gradients would have to be calculated with a finite-difference formula using an automatically chosen step size large enough to avoid gradient explosion. The \textsc{Adam} algorithm already automatically determines the learning rate for each parameter, based on previous results. Therefore, it seems plausible that the step size for the finite-difference computation of the gradient can be treated similarly. Future research could explore this possibility.
\printbibliography
\end{document}